\documentclass{emulateapj}
\usepackage{rotating}

\shorttitle{The Protostar 2MASS J17112318-2724315}
\shortauthors{Riaz et al. }

\begin{document}

\title{2MASS J17112318-2724315: A Deeply-Embedded Low-Mass Protostellar System in the B59 Molecular Cloud}

\author{Riaz, B.\altaffilmark{1} , Mart\'{i}n, E. L.\altaffilmark{1,2} , Bouy, H.\altaffilmark{1} , Tata, R.\altaffilmark{2}  }
\altaffiltext{1}{Instituto de Astrof'sica de Canarias, E38205 La Laguna, Tenerife, Spain}
\altaffiltext{2}{University of Central Florida, Department of Physics, P.O. Box 162385, Orlando, FL 32816-2385}
\email{basmah@iac.es}

\begin{abstract}
We present near-infrared observations of the low-mass deeply-embedded Class 0/I system 2MASS J17112318-2724315 (2M171123) in the B59 molecular cloud. Bright scattered light nebulosity is observed towards this source in the $K_{s}$ images, that seems to trace the edges of an outflow cavity. We report the detection of a low-luminosity protostar 2M17112255-27243448 (2M17112255) that lies $\sim$8$\arcsec$ ($\sim$1000 AU) from 2M171123. This is a Class I system, as indicated by its 2-8 $\micron$ slope and IRAC colors, with an estimated internal luminosity of $\sim$0.3$L_{\sun}$. We estimate a mass of $\sim$0.12-0.25 $M_{\sun}$ for this source, at an age of 0.1-1Myr. Also presented is detailed modeling of the 2M171123 system. The best-fit parameters indicate a large envelope density of the order of $\sim$$10^{-13} g cm^{-3}$, and an intermediate inclination between 53$\degr$ and 59$\degr$. The observed $K_{s}$-band variability for this system could be explained by slight variability in the mass infall rate between 2.5E-5 and 1.8E-5 $M_{\sun}$/yr. The protostar 2M171123 exhibits a rarely observed absorption feature near 11.3 $\micron$ within its 10 $\micron$ silicate band. We find a strong correlation between the strength in this 11.3 $\micron$ `edge' and the $H_{2}O$-ice column density, indicating the origin of this feature in the thickness of the ice mantle over the silicate grains. 

\end{abstract}


\section{Introduction}

The collapse of the cold molecular cloud from which stars are born requires the presence
of rotationally supported disks in order to conserve the angular momentum. On a
timescale of $10^{5}$ yrs, a central protostar is formed surrounded by an infalling envelope
and an accreting disk (Hartmann 2000). Then begins the main accretion or T Tauri phase, during which the central protostar builds up its mass from the surrounding infalling envelope and accretion disk. As gas accretes onto the star, an accretion-driven stellar wind develops that sweeps up infalling material from the surrounding cloud and drives a molecular outflow. The cavities carved out in the circumstellar envelopes by such molecular outflows can be observed in scattered light nebulae. Several surveys based on near-infrared (NIR) and {\it Spitzer}/IRAC observations of outflow cavities in scattered light have enabled detailed study of the morphology of protostellar systems, such as the extent of wind-created cavities, the large-scale geometry and inner structures of the infalling envelopes, as well as the size and structure of the disks (e.g., Padgett et al. 1999; Burrows et al. 1996; Tobin et al. 2008). 

This paper discusses NIR observations and modeling of the deeply embedded low-mass object 2MASS J17112317-2724315 (hereafter 2M171123) in the B59 molecular cloud. B59 is an irregularly shaped nearby ({\it d} = $130^{+30}_{-20}$ pc; Lombardi et al. 2006) dark cloud and is the only known star-forming region in the Pipe Nebula. A recent {\it Spitzer} investigation of the dark cloud core under the ``Cores to Disks Legacy Survey'' (c2d) resulted in doubling the previously known population, with 20 candidate low-mass young stars identified near the core, including two deeply embedded protostellar objects (Brooke et al. 2007; hereafter B07). 2M171123 was classified as a Class 0/I system with a bolometric luminosity of 2.2$\pm$0.3$L_{\sun}$, based on its spectral energy distribution (SED) from NIR through millimeter wavelengths. Extended structures possibly tracing the edges of an outflow cavity were observed towards this protostar in the IRAC 3.6 and 4.5 $\micron$ images (B07). The apparent axis of the northeast extension seemed to be aligned with some of the CO outflow peaks earlier detected by Onishi et al. (1999; hereafter O99), suggesting 2M171123 to be the driving source of this outflow. Considering that no extended emission was observed in the 2MASS images, it was essential to obtain deeper NIR observations of this object to detect scattered light nebulae and conduct a detailed study of the outflow cavity. In $\S\ref{obs}$, we present our NIR observations as well as some archival data utilized for this study, $\S\ref{morphology}$ discusses the morphology of this system as observed in the NIR images along with the detection of a new low-mass object ($\S\ref{faint}$), detailed modeling describing the structure of 2M171123 is presented in $\S\ref{sys_model}$, while $\S\ref{ices}$ attempts to explain the 11.3 $\micron$ `edge' observed in the 10 $\micron$ spectrum for 2M171123, thought to be indicative of the presence of crystalline silicates.

\section{Observations}
\label{obs}

\begin{figure*}
\epsscale{0.9}
\plottwo{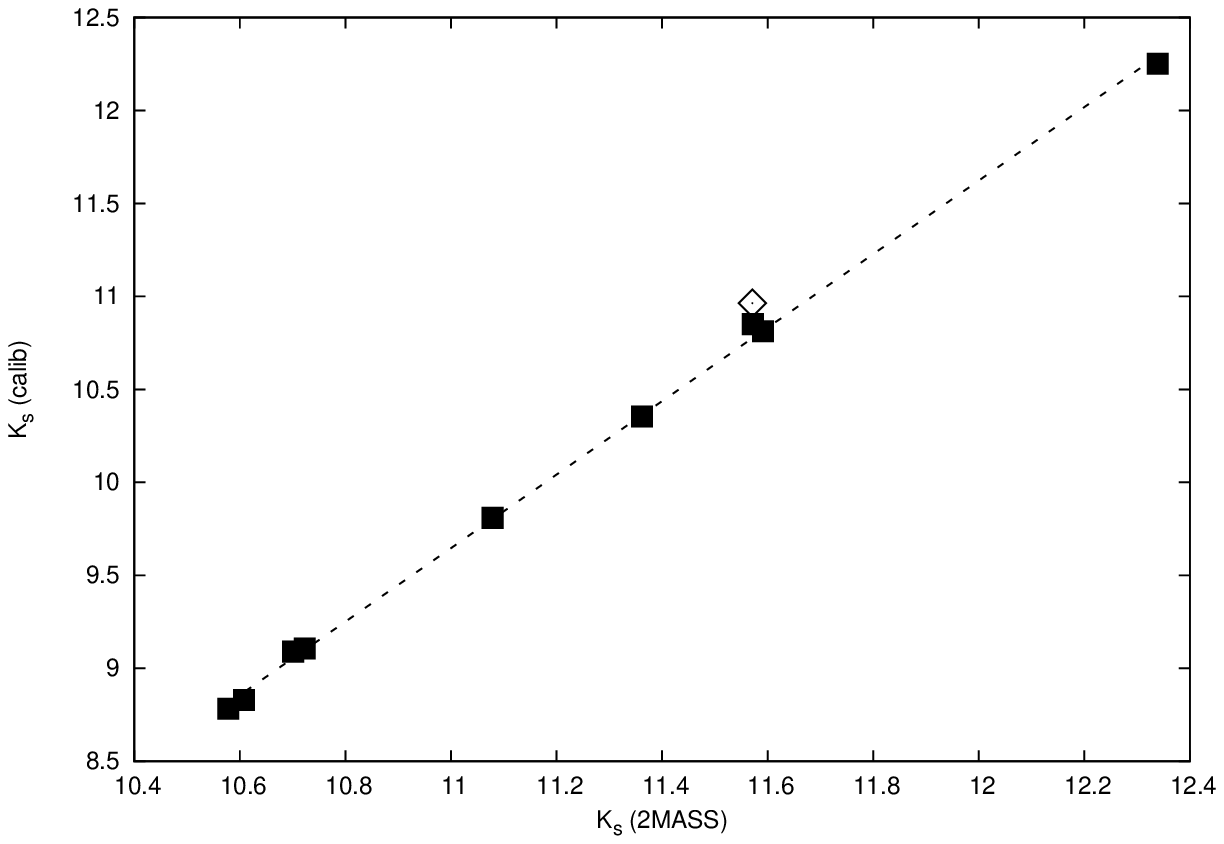}{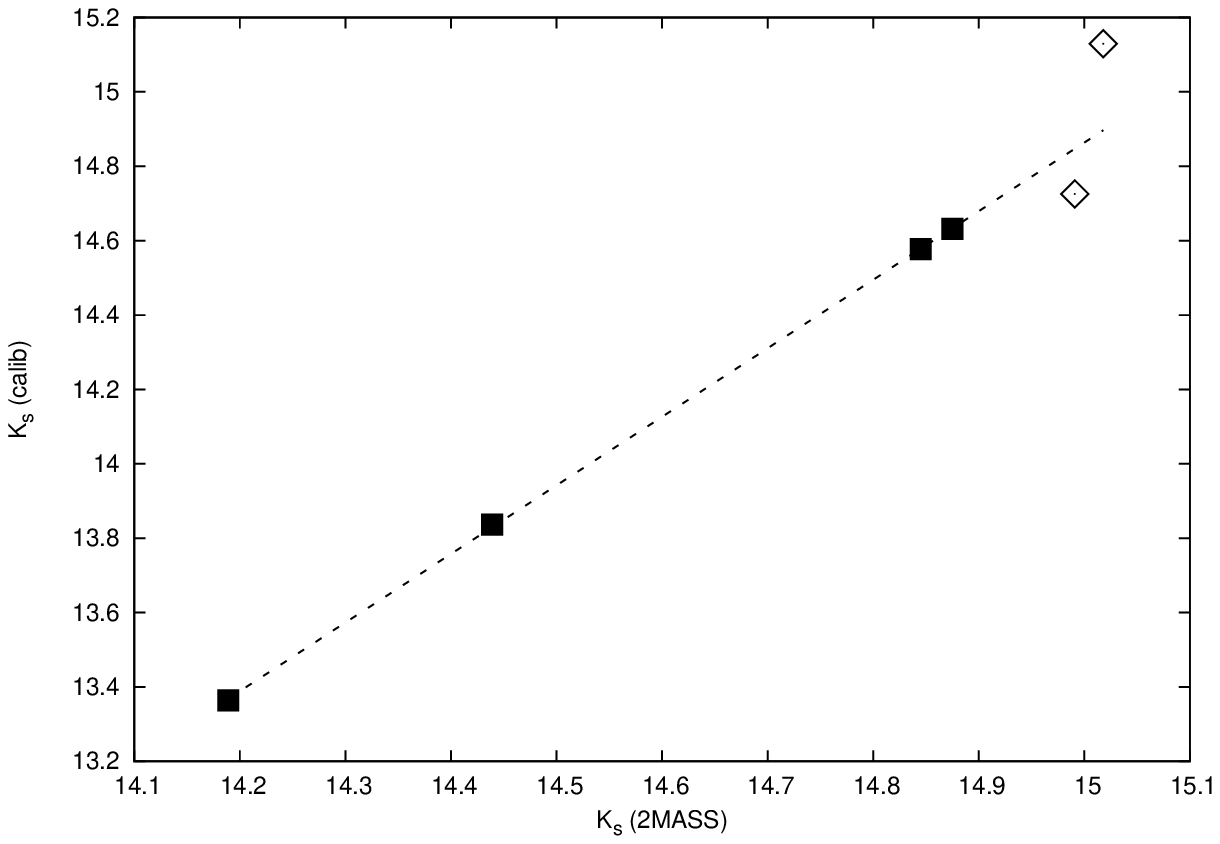}
    \caption{2MASS/$K_{s}$ versus the calibrated $K_{s}$-band photometry for the CTIO ({\it left}) and VLT ({\it right}) data. Dotted line shows a linear least-square fit to the data. Open triangles represent the stars not considered for calibration. }
    \label{calib}
 \end{figure*}

Observations in the {\it J} and $K_{s}$ bands of B59 cloud core were obtained at the CTIO Blanco 4 m telescope on 2008 March 16, using the Infrared Side Port Imager (ISPI). This infrared camera has a field of view of $10.25\arcmin\times10.25\arcmin$, with a pixel scale of 0.3$\arcsec$ pixel$^{-1}$. One pointing was defined at roughly $\alpha = 17^{h}11^{m}, \delta = -27\degr24\arcmin$, with ten dither positions obtained to minimize the effects of cosmic rays and bad CCD columns. The size of the dithers were between 20$\arcsec$ and 30$\arcsec$. The exposure times were 40s and 20s in the {\it J} and $K_{s}$ bands, respectively, with 3 number of coadds obtained, thus resulting in a total time per image of 120s and 60s. Also presented here are {\it H} and $K_{s}$ band archival observations of Pipe nebula obtained at the VLT on 2002 July 28 and 30 [PID 69.C-0426(A)], using the Infrared Spectrometer and Array Camera (ISAAC). The SW imaging mode for ISAAC provides a field of view of $152\arcsec\times152\arcsec$, with a pixel scale of 0.148$\arcsec$ pixel$^{-1}$. The total integration times were 60s and 30s in the {\it H} and $K_{s}$ bands, respectively, with 10 dither positions obtained in each band. The dither sizes were between 2$\arcsec$ and 10$\arcsec$ in the {\it H}-band, and between 6$\arcsec$ and 20$\arcsec$ in the $K_{s}$-band. The raw frames were bias-corrected and flat-fielded using basic IRAF/{\it ccdred} routines. For each object frame, a sky frame was constructed by median-combining all of the dithered observations excluding that object frame, without making any corrections for the offsets. The final sky-subtracted calibrated frames were first aligned to a common $x$ and $y$ position using the IRAF task {\em imshift}, and then averaged using the task {\em imcombine}. Aperture photometry was performed on the combined image for each band using the IRAF task {\em apphot} under the {\em digiphot} package. The data were calibrated using 2MASS magnitudes for bright sources in the same fields. The radial profiles for these bright sources were checked for any saturation effects. The magnitudes of the 2MASS sources that were used for calibration of the CTIO data ranged between {\it J}$\sim$11.5-11.9 mag and $K_{s} \sim$10.5-12.5 mag. For the VLT data, the magnitude range was between 13.6 and 15.7 mag in the {\it H}-band, and between $\sim$14 and 15 mag in the $K_{s}$-band. The calibration uncertainty is estimated to be 0.08 mag and 0.04 mag in the ISAAC/{\it H} and $K_{s}$ bands, respectively, while it is 0.1 mag and 0.07 mag in the CTIO/{\it J} and $K_{s}$ bands. In Fig.~\ref{calib}, we show plots of our calibrated $K_{s}$ versus the 2MASS/$K_{s}$ photometry for the stars used for calibration. The dotted line in these plots indicates a linear least-square fit to the points, and we find a good correlation with the 2MASS/$K_{s}$ photometry for both the VLT and CTIO data. Stars that lie offset from the linear fit are represented in Fig.~\ref{calib} by open triangles, and have not been used as calibrators. These sources are 2MASS J17112508-2724425 (2MASS/$K_{s}$=11.57) in the CTIO data, and 2MASS J17112111-2725004  (2MASS)/$K_{s}$=14.99) and 2MASS J17112749-2726029 (2MASS/$K_{s}$=15.02)  in the ISAAC observations. As mentioned, the radial profiles for these objects were checked for any saturation effects, and the linear fit shown in Fig.~\ref{calib} indicates that saturation should not have affected the calibration for both the ISAAC and CTIO data. Table 1 lists the photometry thus obtained. 2M171123 could not be detected in the {\it J}-band, and an upper limit has been determined. We also list in Table 1 the 2MASS measurements for this target. In comparison with 2MASS/$K_{s}$, the CTIO measurement is $\sim$1.2 magnitude brighter, while a less significant brightening of $\sim$0.4 mag is observed in the ISAAC observations. \S\ref{sys_model} discusses the possible causes for the observed variability in the $K_{s}$-band. $\S\ref{ices}$ compares the {\it Spitzer}/IRS spectra for 2M171123 with other protostellar systems. The spectra were obtained from the {\it Spitzer} archives (PID 172, 179 and 20604). The spectra were extracted and calibrated using the Spitzer IRS Custom Extraction (SPICE) software provided by the {\it Spitzer Science Center}. For the purpose of modeling the 2M171123 system, we have utilized {\it Spitzer}/IRAC and MIPS photometry from B07, as well as submillimeter and millimeter observations from Reipurth et al. (1996) and Wu et al. (2007).

\begin{deluxetable*}{ccccccccccccc}
\tabletypesize{\tiny}
\tablewidth{0pt}
\tablecaption{Observations}
\tablehead{
\colhead{Star} & $\alpha$ (J2000)&$\delta$ (J2000)& & \colhead{{\it J}} & \colhead{{\it H}} & \colhead{$K_{s}$} & \colhead{[3.6]} &
\colhead{[4.5]} & \colhead{[5.8]} & \colhead{[8]} & \colhead{[24]} \\
}
\startdata
2M171123 &17 11 23.18 & -27 24 31.5 &CTIO &19.63\tablenotemark{a} &  & 13.87$\pm$0.11 & 11.59$\pm$0.03 & 9.30$\pm$0.03& 7.87$\pm$0.01& 6.50$\pm$0.01&0.47$\pm$0.2 \\
		  &&&ISAAC& 				& 19.47$\pm$0.16&15.46$\pm$0.07& &&&\\		  
		  &&&2MASS &18.80\tablenotemark{a}&17.80\tablenotemark{a}&15.08$\pm$0.14&&&&\\
\hline		  
		  
2M17112255 &17 11 22.55 & -27 24 34.38 &CTIO & 20.62\tablenotemark{a} & & 15.98$\pm$0.14 &14.2$\pm$0.10&13.1$\pm$0.1&12.7$\pm$0.2&11.9$\pm$0.1&3.66$\pm$0.2\\
		&&& ISAAC && 20.51$\pm$0.3 & 16.74$\pm$0.10& &&&\\
		&&&2MASS&&&&&&& \\
                     
\enddata
\tablenotetext{a}{Upper limit.}
\end{deluxetable*}

\section{Observed Morphology}
\label{morphology}
 
\begin{figure*}
\plotone{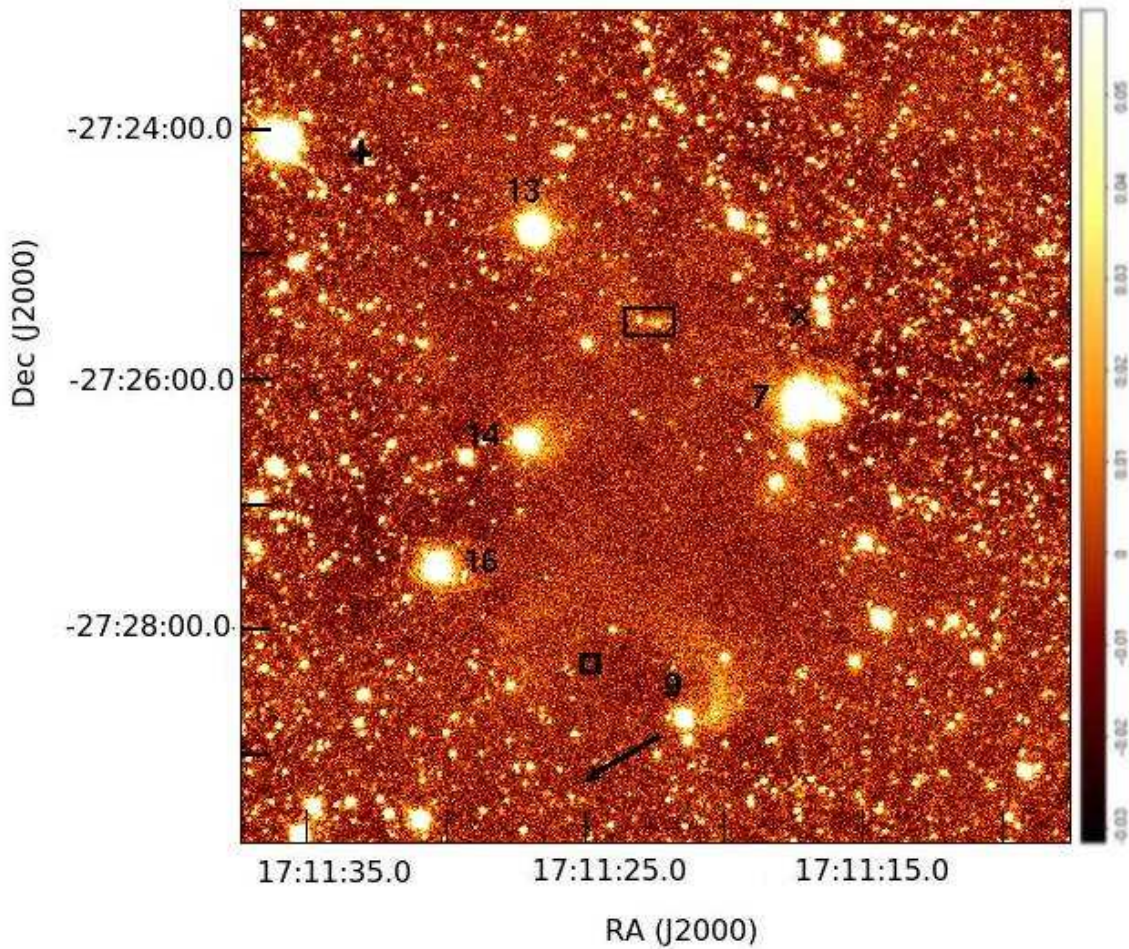}
 \caption{False color CTIO/$K_{s}$ image, roughly centered around 2M171123 (marked by a rectangle). Also marked are the locations of $C^{18}O$ peak (open square), red ouflow peak (`x'), and the blue peaks (plus signs). Also indicated are some of the sources from B07 discussed in $\S\ref{morphology}$. The arrow points towards the third blue outflow peak that lies $\sim$0.08 pc southeast from Source 9. The color intensity scale in units of MJy $sr^{-1}$ is shown on the right. The image is 400$\arcsec$ on a side, which corresponds to 0.25 pc at a distance of 130 pc.  }
 \label{morph1}
\end{figure*}

O99 detected 14 $C^{18}O$ cores in the B59 region. One of the cores (named Core 1 in their paper) was found to have an emission intensity a factor of $\sim$2 larger than all other cores. Two sources were found to be within 0.1 pc of this $C^{18}O$ core: 2M171123 and IRAS 17081-2721 (2MASS J17111726-2725081; source 7 in B07). The CO outflow detected was also found to be closely associated with the $C^{18}O$ peak (within 0.3 pc). The close proximity of 2M171123 and IRAS 17081-2721 thus suggested that one or both of them could be the driving sources of this outflow. O99 interpreted the outflow detected as a possible superposiion of two outflows; one outflow has a position angle of $\sim$20$\degr$ east of north, consisting of an elongated blue lobe $\sim$0.7 pc and a compact red lobe within $\sim$0.3 pc of the $C^{18}O$ core. The other outflow has a position angle of $\sim$ -40$\degr$ and consists of a blue and a red lobe with spatial extents of 0.4-0.5 pc. Figure~\ref{morph1} shows the CTIO/$K_{s}$ image, roughly centered around the 2M171123 location. Also marked are the locations of the $C^{18}O$ peak emission, the CO red and blue outflow peaks, as well as some of the sources from B07 discussed in this section. 

\begin{figure*}
\plottwo{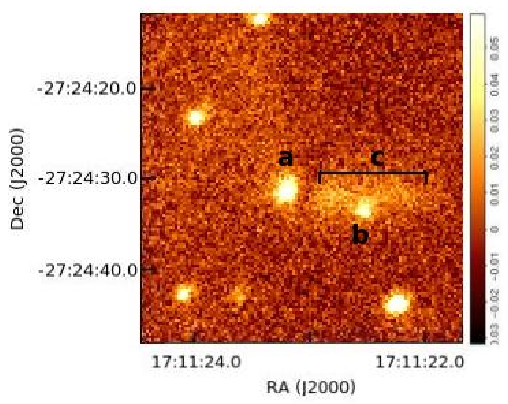}{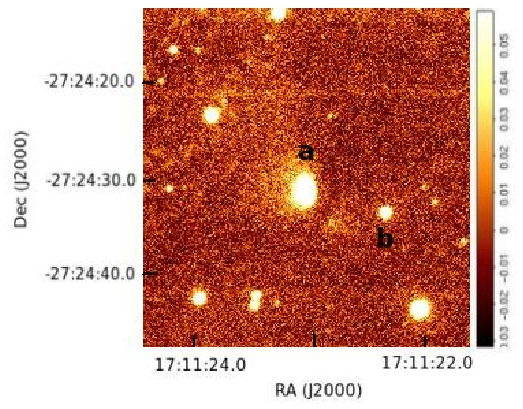}
\epsscale{0.4}
\plotone{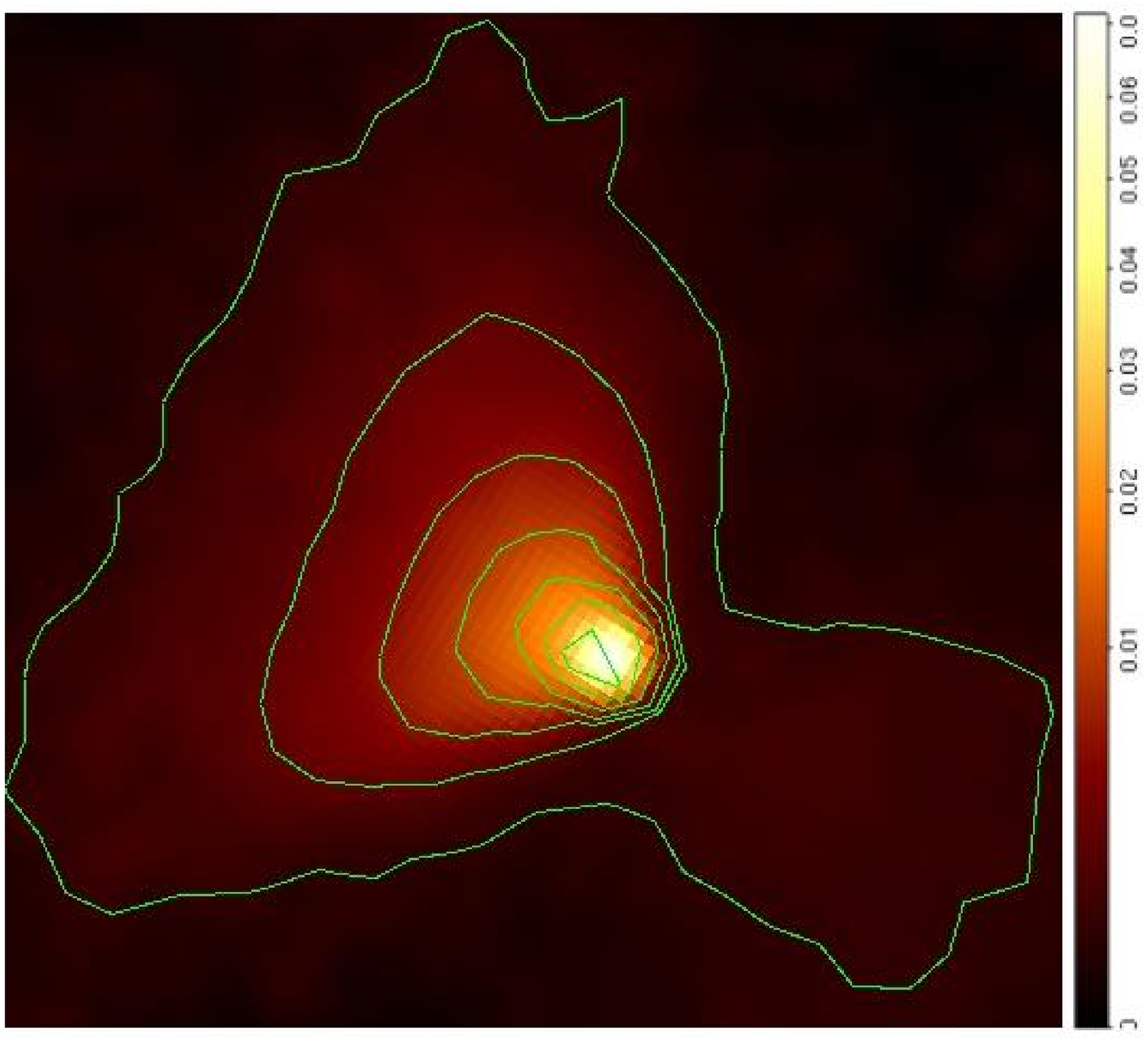}
    \caption{False color images of 2M171123 in the CTIO/$K_{s}$-band ({\it top, left}), and the ISAAC/$K_{s}$-band ({\it top, right}). The labels {\it a, b and c} in the CTIO image indicate 2M171123, 2M17112255, and the faint bar-like extension, respectively. The color intensity scale in units of MJy $sr^{-1}$ is shown on the right of each image. The images are 35$\arcsec$ on a side, which corresponds to 0.02 pc at a distance of 130 pc. The bottom panel shows a $K_{s}$-band  (2.16 $\micron$) model image for 2M171123. The image is 25$\arcsec$ on a side, corresponding to 0.016 pc at 130 pc. The overplotted contours correspond to intensity levels of 0.0006, 0.002, 0.005, 0.01, 0.015, 0.02 and 0.035 MJy $sr^{-1}$. }
    \label{morph2}
 \end{figure*}

We observe a curved extended structure near 2M171123 that extends to roughly 0.5 pc along the northeast direction. The extension is also clearly seen in the ISAAC/$K_{s}$ observations (Fig.~\ref{morph2}). The apparent axis of this structure seems to be aligned with the first outflow discussed by O99, i.e. at a position angle of 20$\degr$ east of north, and may trace the edges of this outflow cavity. This was first noted by B07 in their {\it Spitzer}/IRAC observations, who gave a larger position angle of $\sim$50$\degr$ east of north. The boundaries of the eastern cavity as observed in the IRAC 3.6 and 4.5 $\micron$ channels by B07 however seem more constraining for the position angle than the northeast nebula observed in our $K_{s}$ observations.  

Another extension is observed to the west of 2M171123 in the CTIO/$K_{s}$ image (Fig.~\ref{morph2}a; marked by `c'), with a spatial extent of about 0.007 pc. This faint emission could be due to the decreased sensitivity and worse seeing ($\sim$0.8-1$\arcsec$) in the CTIO data, as it is undetected in the more sensitive ISAAC/$K_{s}$ image (Fig.~\ref{morph2}b). It is also possible that the faint extension seen in the CTIO/$K_{s}$ is the northern boundary of the southwestern cavity, since this emission is at the expected location of that boundary (based on reflecting the southern boundary of the northeastern cavity). The location of this structure is quite consistent with what is observed in the IRAC channels, although B07 have noted this extension to be confused by a superposed background source. This faint source (2M17112255-27243448; marked by `b') is observed very clearly in the ISAAC/$K_{s}$ observations (Fig.~\ref{morph2}b), and we discuss it in more detail in $\S\ref{faint}$. 

A more prominent curved structure is observed in the CTIO/$K_{s}$-band towards the star 2MASS J17112153-2727417 (source 9 in B07) that seems to extend in a counter-clockwise direction (from bright to faint) first towards the west, then northwest (Fig.~\ref{morph1}). This structure may be associated with a third blue outflow peak (marked by an arrow) that lies off the image towards the southeast. However the apparent axis of this structure does not seem to align with this peak. The intensity of this blue lobe is weaker by a factor of $\sim$2 compared to the red lobe, and lies outside the edge of the $C^{18}O$ core (O99). Source 9 lies $\sim$0.08 pc from this blue peak, but its closer proximity to the $C^{18}O$ peak ($\sim$0.03 pc) suggests that this strong extension could be due to cloudshine emission.

No clear extended emission is observed towards the red peak or the blue outflow peak west of 2M171123, suggesting that the second outflow at a position angle of -40$\degr$ may be too weak in intensity to produce any strong scattered emission. Some strong nebulosity is observed towards sources 7, 13 and 14. A fainter structure of small spatial extent ($\sim$0.02 pc) was noted by B07 southeast of 2M171123, that may be the southern or the southeastern boundary of the northeastern lobe. It seems that the southern boundary of the $K_{s}$-band nebulosity is quite consistent with what has been observed by B07 in the IRAC bands. This structure however is not very prominent in the $K_{s}$ observations.

\subsection{Observed colors for 2M17112255}
\label{faint}

\begin{figure*}
\plottwo{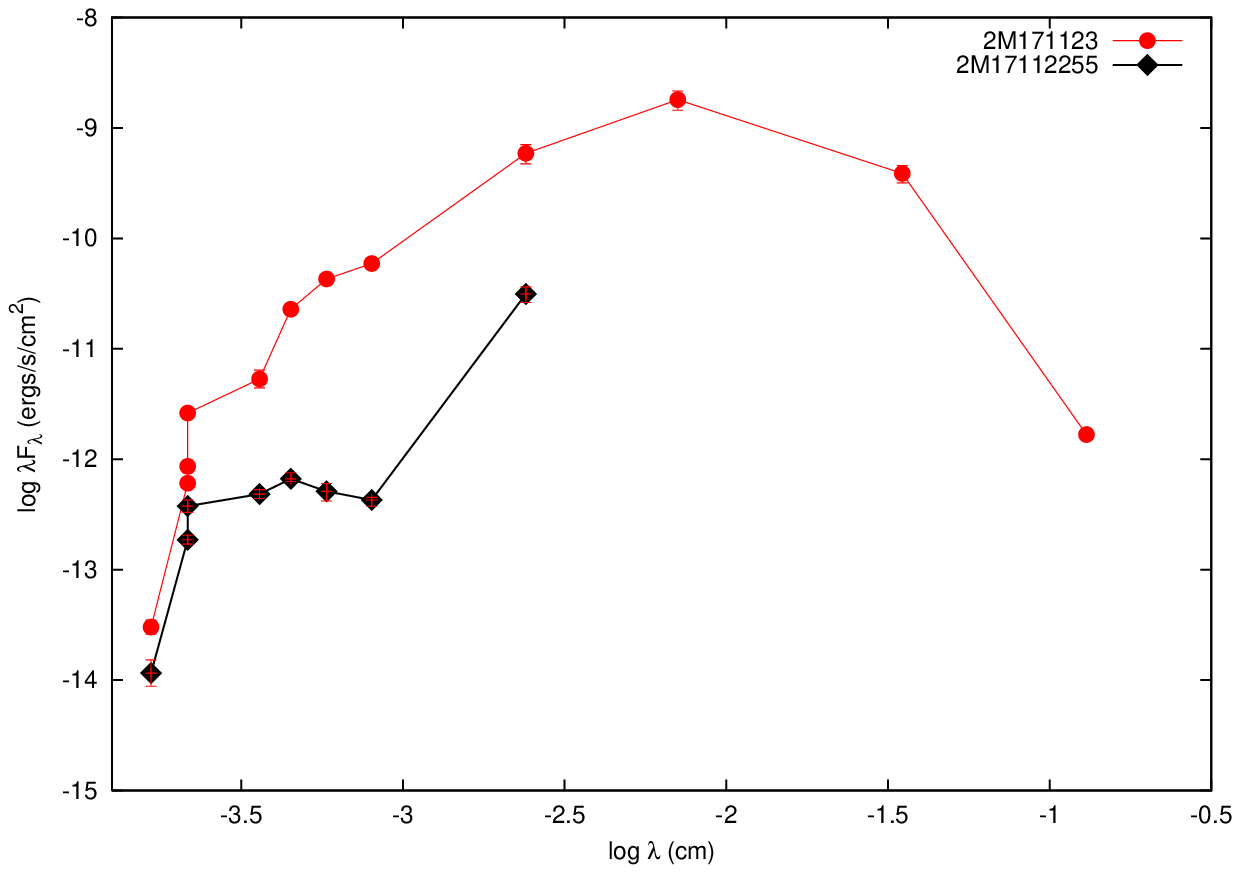}{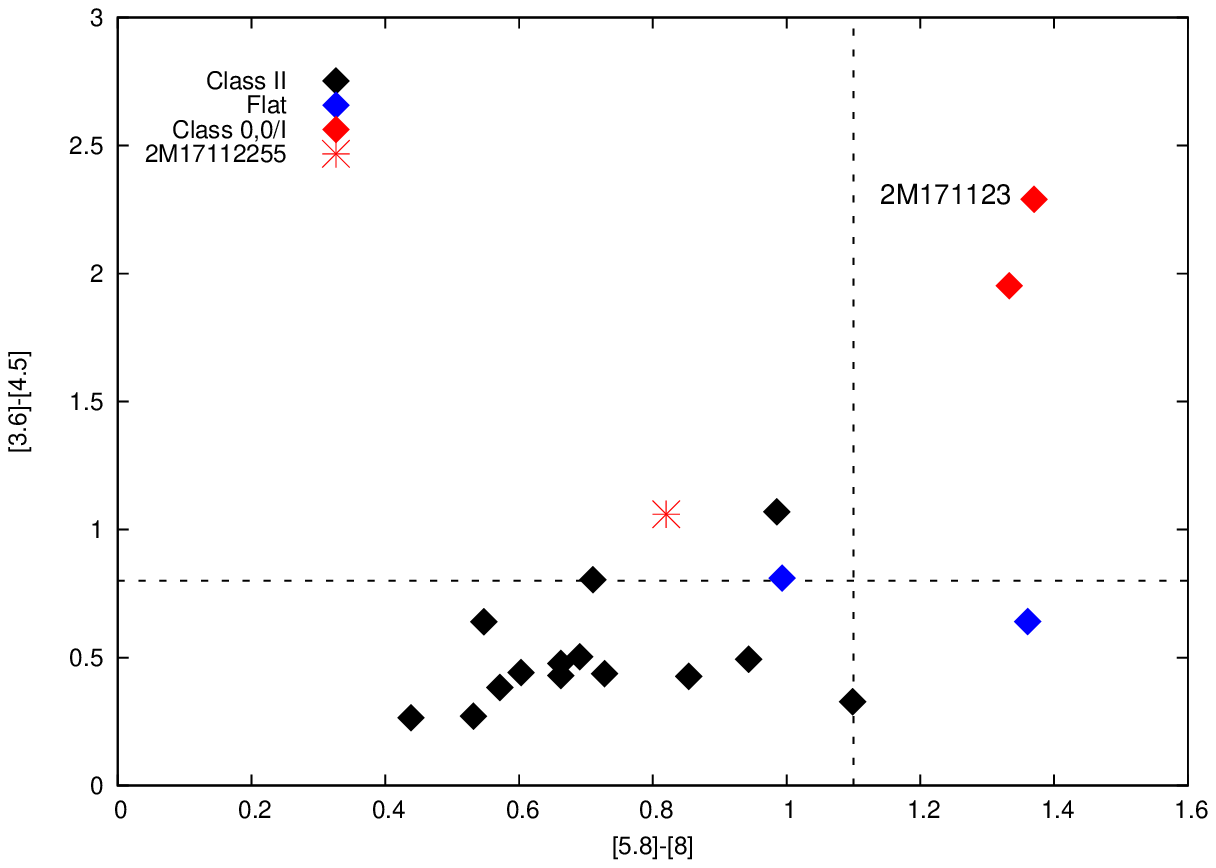}
\plottwo{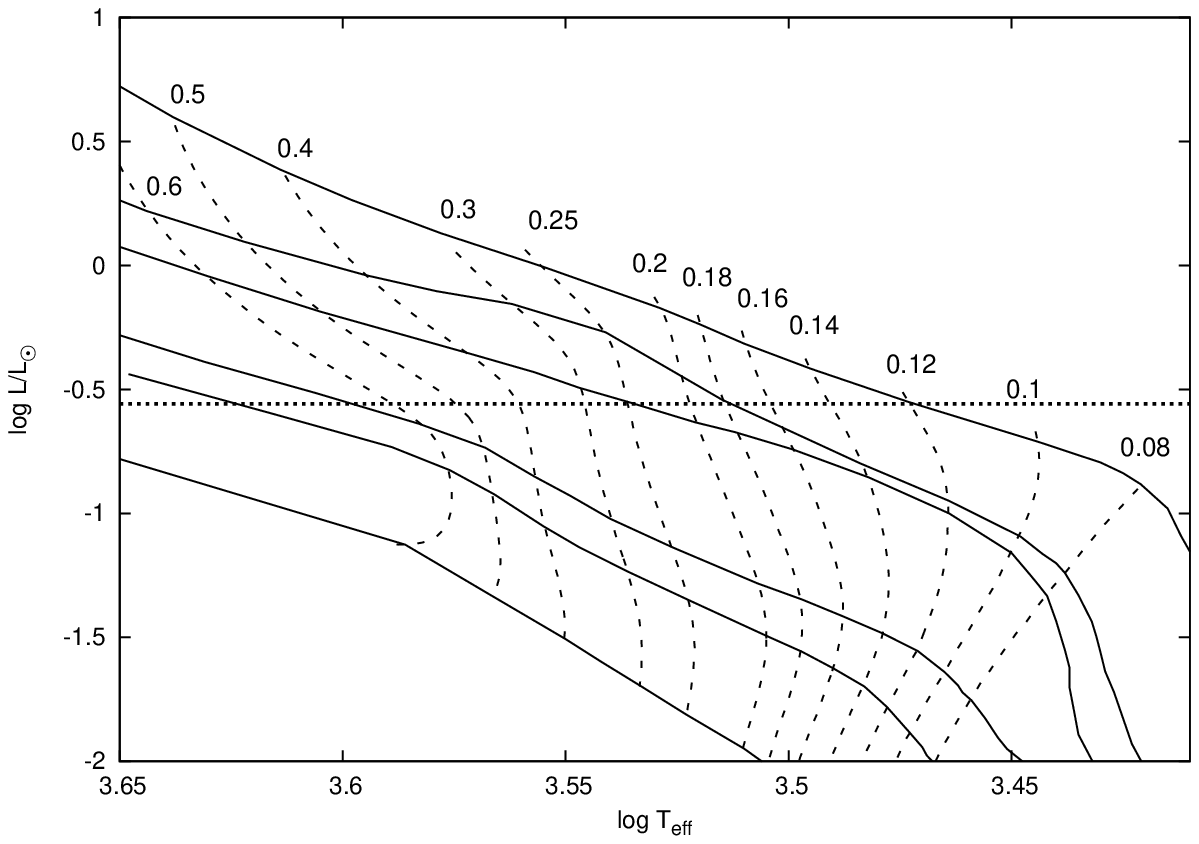}{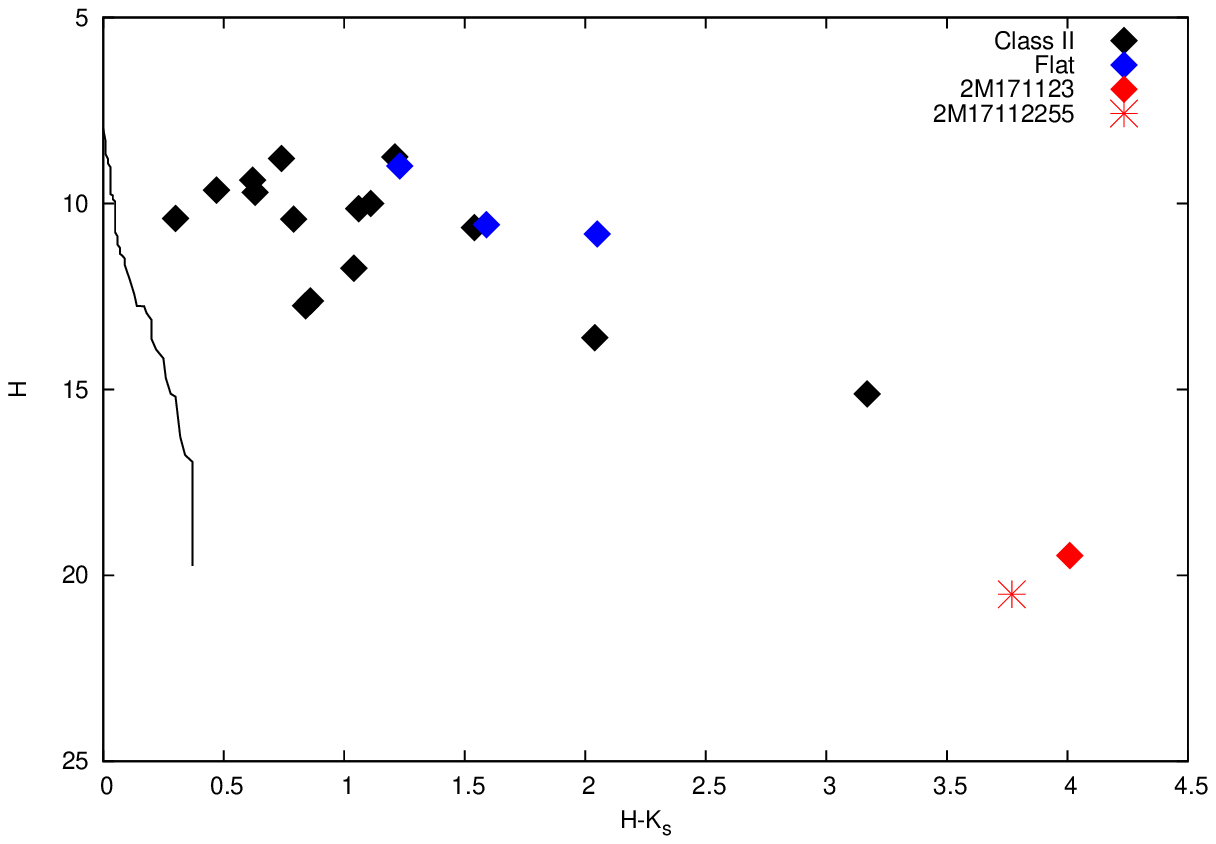}
    \caption{{\it Top, left} (a): SEDs for 2M171123 and 2M17112255. {\it Top, right} (b): IRAC color-color diagram for the B59 objects. {\it Bottom, left} (c): Evolutionary models from D'Antona \& Mazzitelli (1998). The masses in $M_{\sun}$ are marked. The isochrones (from top to bottom) are for 0.1 Myr, 0.5 Myr, 1 Myr, 10 Myr and 50 Myr. The horizontal dotted line indicates the internal luminosity for 2M17112255. {\it Bottom, right} (d): NIR color-magnitude diagram for the B59 objects. Bold line indicates the zero-age main sequence.   }
    \label{ccd}
 \end{figure*}

In the CTIO and ISAAC images, we observe a faint source ($\alpha = 17^{h}11^{m}22.55^{s}, \delta = -27\degr24\arcmin34.38\arcsec$; hereafter 2M17112255), $\sim$8$\arcsec$ ($\sim$1000 AU) westward from 2M171123. No 2MASS counterpart for this object could be found, and it was undetected in our {\it J}-band observations. The NIR photometry along with a {\it J}-band upper limit is listed in Table 1. This is also a deeply-embedded object, with an $A_{v}\sim$ 19 - 30 estimated from the extinction maps in B07. The c2d catalog lists a source, identified as SSTc2d J171122.5-272434, at the position of 2M17112255. The `object\_type' for this source is given as `YSOc\_red', i.e. a candidate YSO with a MIPS 24 $\micron$ to IRAC flux ratio greater than 3. The c2d IRAC and MIPS flux densities are listed in Table 1. As discussed in the c2d Legacy Project report\footnote{available at http://peggysue.as.utexas.edu/SIRTF/}, to produce an IRAC-MIPS1 band-merged list, a 4.0$\arcsec$ matching radius is utilized, while to merge this list with the MIPS2 (70 $\micron$ channel) source list, a larger radius of 8.0$\arcsec$ is considered. The 24 $\micron$ photometry for 2M17112255 should thus have no contribution from 2M171123. There is no MIPS2 detection listed in the c2d catalog for 2M17112255. The MIPS2 detection quality for 2M171123 is listed as `P', indicating that there were multiple catalog detections near this source, which is expected given the 8$\arcsec$ matching radius. The 70 $\micron$ photometry is thus for the composite source. Fig~\ref{ccd}a shows the SEDs for the two objects. The 2 to 8 $\micron$ slope of the SED is measured to be 3.67 for 2M171123, and 0.64 for 2M17112255. Following Wilking et al. (2001), we define the spectral index as  $\alpha$ = {\it d} log ($\lambda F_{\lambda}$)/ {\it d} log $\lambda$, where $\alpha >$ 0.3 is a Class I source, 0.3 to -0.3 is a `Flat' source, -0.3 to -1.6 is a Class II source, and Class III sources have $\alpha <$ -1.6. Employing this criteria classifies both objects as Class I sources, as already noted for 2M171123 by B07. Fig.~\ref{ccd}b compares the IRAC color-color diagram for the 20 low-mass candidate members in B59 with 2M17112255. The catalogued c2d photometry of 2M17112255 yields [5.8]-[8.0] = 0.82$\pm$ 0.2, and [3.6]-[4.5]=1.06$\pm$0.14. In their study of IRAC colors of young stellar objects, Allen et al. (2004) have considered the boundary between Class I and II objects to lie approximately at [3.6]-[4.5] $\sim$0.8 and [5.8]-[8] $\sim$ 1.1. Exceptions to this criteria are low-luminosity Class I objects that would show bluer [5.8]-[8] colors at a given [3.6]-[4.5] color. Furthermore, Class I objects with low envelope densities of the order of $\sim 10^{-14} g cm^{-3}$ show bluer [3.6]-[4.5] colors at a given [5.8]-[8] color (Allen et al. 2004; Fig. 1). In comparison with the envelope models and the criteria for Class I objects discussed in Allen et al., the location of 2M17112255 in Fig.~\ref{ccd}b indicates that it is consistent with being a low-luminosity ($\sim$0.1$L_{\sun}$) Class I object with a low envelope density of $\sim 3E-14 g cm^{-3}$. 

For such embedded protostars, however, it is important to determine the internal luminosity, $L_{int}$, of the source, i.e., the total luminosity of the central protostar that excludes any emission from the external heating of the circumstellar envelope by the interstellar radiation field (e.g., Dunham et al. 2008). An estimate on $L_{int}$ could be obtained via 1-D or 2-D radiative transfer modeling of the observed SED. A detection at 70 $\micron$ though is crucial for such modeling, since the observed flux density at this wavelength is least affected by the source geometry and the presence or absence of a disk, and can better constrain the resulting model-fits (Dunham et al. 2008). As mentioned above, the 70$\micron$ photometry listed in Table 1 is for the composite  source i.e. includes emission from both 2M171123 and 2M17112255. Our attempts at modeling the NIR to 24 $\micron$ SED for 2M17112255 resulted in largely degenerate model fits, mainly due to the sudden increase in the flux densities between 8 and 24 $\micron$, and the lack of data at longer wavelengths. We have instead considered the linear least-square relation between the observed flux at 24 $\micron$ and $L_{int}$, obtained by Dunham et al. (2008) for a set of 11 embedded low-luminosity protostars. The relation is derived in the log-log space and is given as:

\begin{equation}
log ~ \lambda F_{\lambda} = 0.87 \pm 0.20 ~(log ~ L_{int}/L_{\sun}) - 10.05 \pm 0.17,
\end{equation}

\noindent where log $\lambda F_{\lambda}$ is the flux at 24 $\micron$ in cgs units ($ergs ~ s^{-1} cm^{-2}$), normalized to 140 pc. Using this relation, $L_{int}$ for 2M17112255 is estimated to be $0.27^{+0.21}_{-0.11} L_{\sun}$. We note that while Dunham et al. have found a tight correlation between the 70$\micron$ flux and $L_{int}$, the correlation at 24 $\micron$ is weak (reduced $\chi^{2}$ value of 85, compared to 3 in the 70 $\micron$ case). Thus the estimate on $L_{int}$ for this source could best be considered as an upper limit. We consider the pre-main sequence evolutionary models by D'Antona \& Mazzitelli (1994; 1998) to estimate the mass for 2M17112255 (Fig.~\ref{ccd}c). The horizontal dotted line in this figure marks the estimated internal luminosity for this source. The age for B59 is not well-defined, and we consider the estimated range of 0.5 - 1 Myr provided by B07. Within this range, the mass for 2M17112255 is estimated to be between $\sim$0.18 and 0.25 $M_{\sun}$. If this object is as young as $\sim$0.1 Myr, which is an age more consistent with it being a protostellar system, then its estimated mass could be as small as $\sim$0.12 $M_{\sun}$. Considering the uncertainty on the internal luminosity, $log L_{int}/L_{\sun}$ could be as small as $\sim$ -0.8, placing 2M17112255 close to the sub-stellar boundary at an age of 0.1 Myr. This therefore could be the first candidate brown dwarf in the B59 molecular cloud, though follow-up spectroscopy would be valuable in confirming its sub-stellar nature. 

We consider Poisson statistics to determine the probability for 2M17112255 to be a background object. The surface density of young stars in the B59 core is found to be $\sim$200 $pc^{-2}$ (B07). This results in the expected number of stars within an 8$\arcsec$ radius to be 0.0051, indicating a weak 0.5\% probability of detecting a background source within 8$\arcsec$ of a B59 member. Among the 20 candidate low-mass stars in B59, there is then a 9 in 10 chance of not finding a background star within this radius. At a separation of $\sim$8$\arcsec$, 2M17112255 thus has a weak 10\% probability of being a background star or a chance alignment. Fig.~\ref{ccd}d compares the NIR colors for 2M17112255 with 20 other candidate young stars in B59 listed in B07. The observed intensity of 2M17112255 in the low-extinction ISAAC/$K_{s}$ image may be considered as an estimate of the intrinsic brightness of this object at this wavelength. In the {\it H}/{\it H-K} color-color diagram, 2M17112255 does not appear as a clear outskirt and follows the trend of redder {\it H-K} colors with fainter {\it H}. The zero-age main sequence (ZAMS) shown in the diagram was assembled from data presented in Bessell (1991) and Leggett et al. (1992). All of the 20 stars including 2M17112255 lie above the ZAMS, as expected for pre-main-sequence stars, with the more evolved Class II systems lying closer to the ZAMS than the less evolved ones categorized as `Flat' and Class 0/I or I sources by B07. The location of 2M17112255 thus suggests that it is unlikely to be a background field star. 2M17112255 and 2M171123 could form a wide binary pair. It is interesting to note that the three known binary systems in B59, B59-1, B59-2 and LkH$\alpha$ 346, have projected separations $\ga$400 AU (Chelli et al. 1995; Reipurth \& Zinnecker 1993; Koresko 2002), suggesting such wide systems to be more probable in this cloud core$\footnote{B59-1 is also known to have a tertiary companion at a projected separation of $\sim$13 AU (Koresko 2002).}$. However, a projected separation of $\sim$1000 AU between 2M171123 and 2M17112255 seems too large to form a stable gravitationally-bound system. The escape velocity for 2M17112255 from the primary is only $\sim$0.6km/s, with a binding energy of $\sim$1E35 J (for a mass of 0.2$M_{\sun}$). A weakly bound system such as this one may still be too young at an age of $\sim$1 Myr to be disrupted, even at such a large separation.

\section{Modeling}
\label{sys_model}

We have used the two-dimensional radiative transfer code by Whitney et al. (2003) to model the 2M171123 protostellar system. The main ingredients of the model are a rotationally flattened infalling envelope, bipolar cavities, and a flared accretion disk in hydrostatic equilibrium. For the circumstellar envelope, the angle-averaged density distribution varies roughly as $\rho \propto r^{-1/2}$ for $r << R_{c}$, and $\rho \propto r^{-3/2}$ for $r >> R_{c}$. Here, $R_{c}$ is the centrifugal radius and is set equal to the disk outer radius. It is the distance from the central protostar along the radial direction where material infalling from the envelope lands onto the disk surface. A smaller value for $R_{c}$ would mean that high-density material would pile up closer to the protostar (e.g., Furlan et al. 2008). Thus increasing the value for this parameter would decrease the optical depth to the center of the envelope. 

The disk density is proportional to $\varpi^{-\alpha}$, where $\varpi$ is the radial coordinate in the disk midplane, and $\alpha$ is the radial density exponent. The disk scale height increases with radius, $h=h_{0}(\varpi / R_{*})^{\beta}$, where $h_{0}$ is the scale height at $R_{*}$ and $\beta$ is the flaring power. The disk density therefore exhibits a Gaussian distribution about the disk midplane, i.e., falls off exponentially with scale height above and below this dense region, as well as in the radial direction towards larger radii. The disk extends from the dust destruction radius, $R_{sub}$ = $R_{*} (T_{sub}/T_{*})^{-2.1}$, to some outer disk radius, $R_{d,max}$. The dust sublimation temperature is adopted to be 1600 K. The disk is truncated sharply at its inner edge, and the region between the inner edge and the star is considered to be either devoid of any disk material or may contain some optically thin dust. 

Bipolar cavities are included in the models to account for the extended scattered emission observed at NIR wavelengths. The cavities extend from the center of the protostar to the outer radius of the envelope. We have adopted the curved cavity shape, the structure of which follows $z = a\varpi^{\beta}$, where $\varpi = (x^{2} + y^{2})^{1/2}$. Here {\it a} is a constant determined by a relation between the envelope radius and the cavity opening angle, and {\it b} is the power of the polynomial defining the cavity shape. The shape parameter determines how quickly the cavity widens in the envelope. A small amount of dust is included in the cavity with constant density, $n_{H_{2}} = 2 \times 10^{4}  cm^{-3}$ (Whitney et al. 2003). 

Table 2 lists the best-fit model parameters for the 2M171123 system. The stellar parameters used are illustrative: adequate for our purposes of modeling but not meant to be definitive. The values listed for these parameters are typical of T Tauri stars in Taurus, and were not varied during the modeling. We have used large grains in the dense disk midplane, with a size distribution that decays exponentially for sizes larger than 50 $\micron$ up to 1 mm. The ISM-like grains with $a_{max} \sim$ 0.25 $\micron$ have been placed in the disk atmosphere and the outflow region. The grain model used in the envelope region is similar in size to the ISM-grains, except includes a layer of water ice on the grains that covers the outer 5\% of the radius. 

\begin{figure}
\plotone{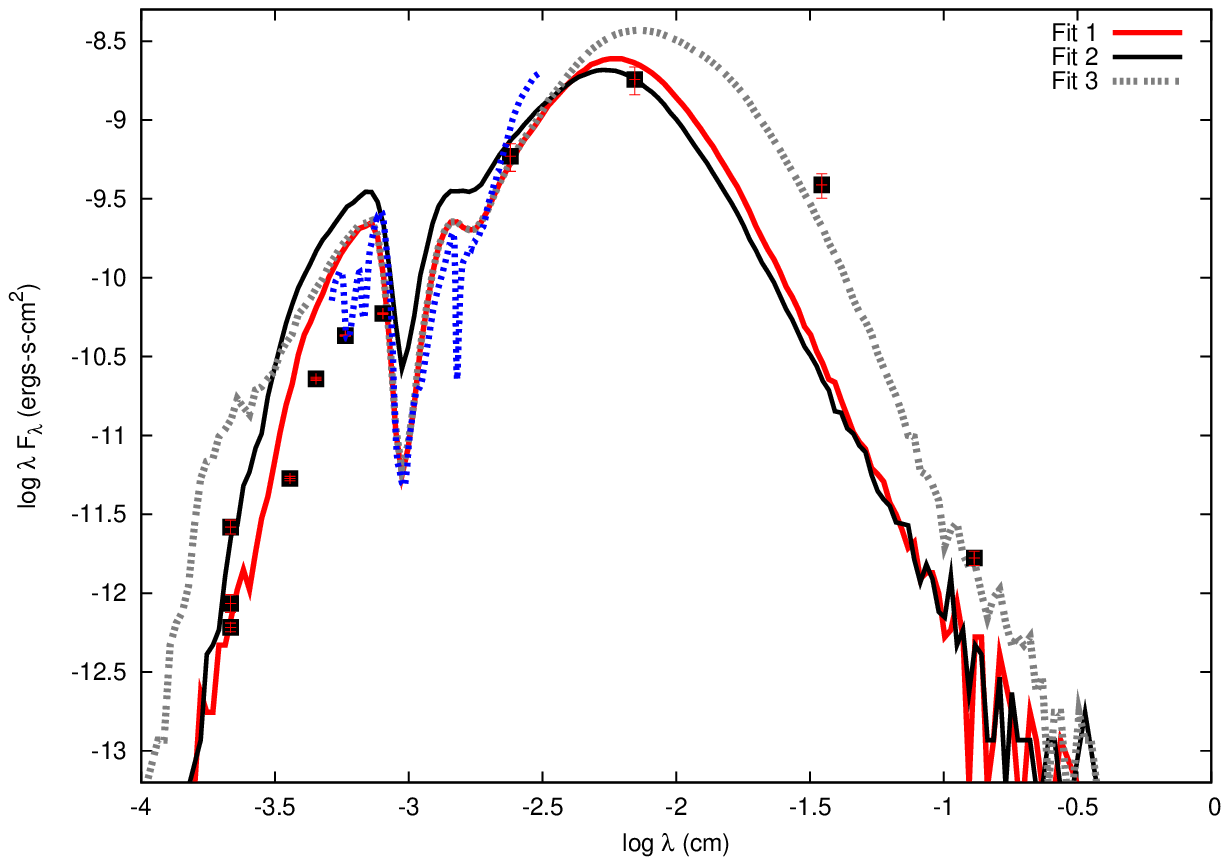}
\plotone{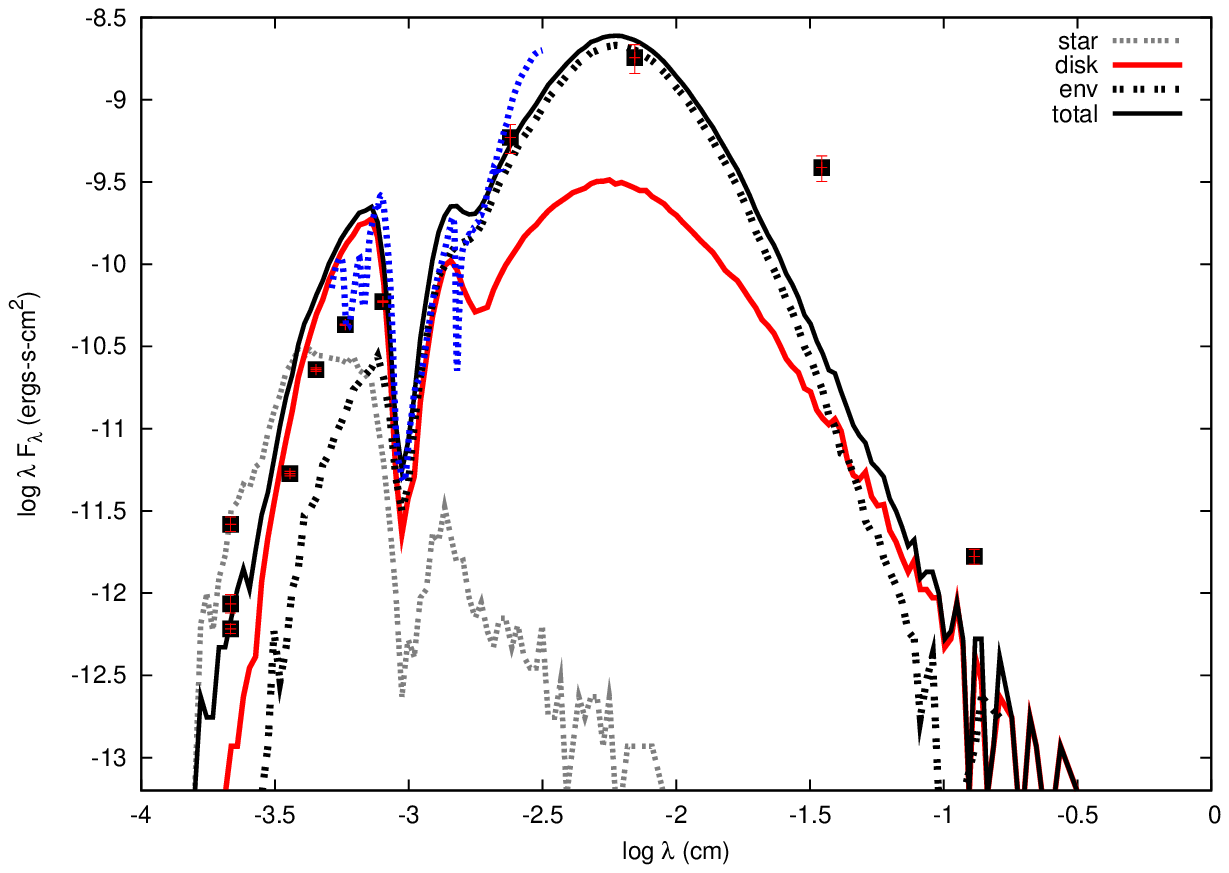}
\plotone{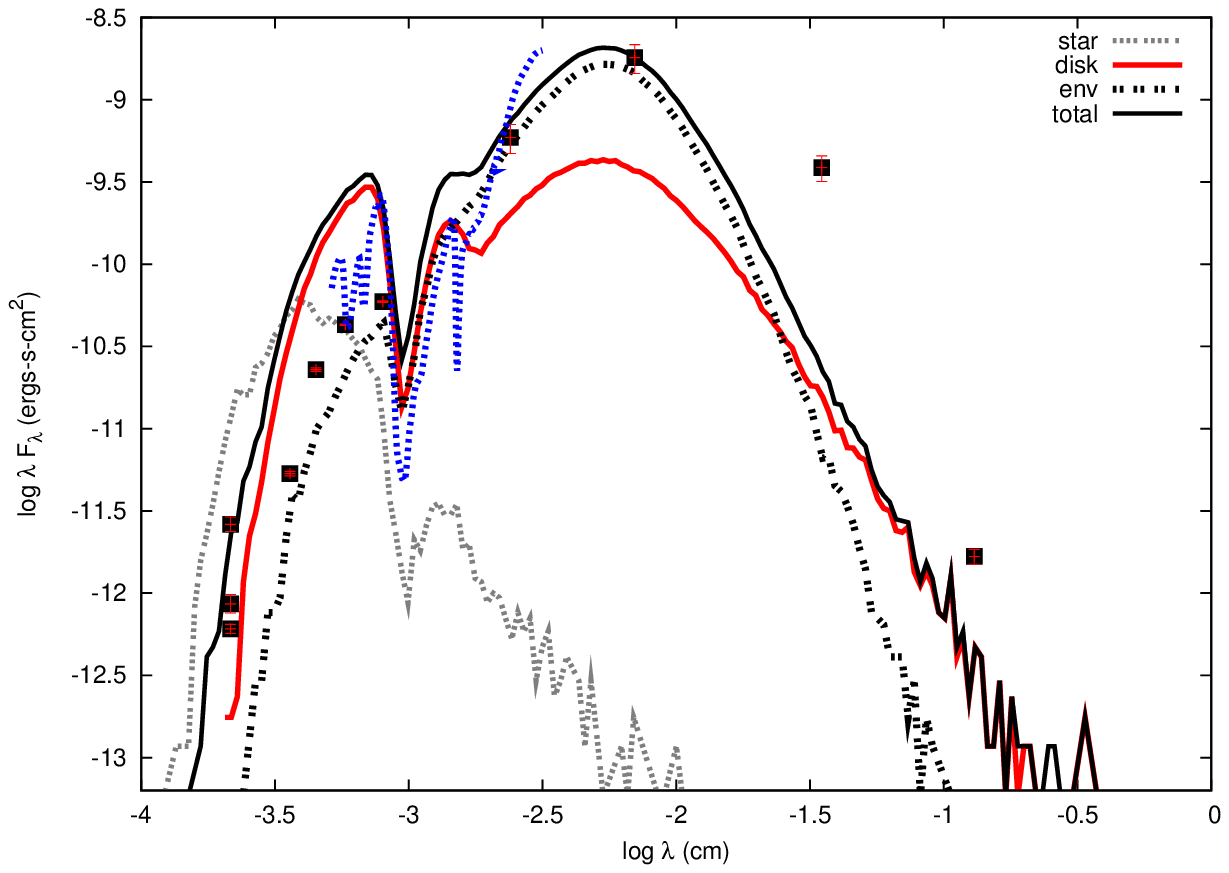}
    \caption{{\it Top, (a)}: The three different fits required to model the 2M171123 system. Separate contributions from the disk, envelope and the star are shown in the middle panel for Fit\#1, and in the bottom panel for Fit\#2. }
    \label{modeling}
 \end{figure}

\begin{deluxetable}{ccccccccccccc}
\tablecaption{Model Parameters}
\label{model}
\tablewidth{0pt}
\tablehead{
\colhead{Parameter} & \colhead{Value} \\ 
}
\startdata
$R_{*}$ & 3.4 $R_{\sun}$ \\
$T_{*}$ & 3300 K\\
$M_{*}$ & 0.28 $M_{\sun}$ \\
$\dot{M}_{env}$ & 1.8E-5 - 2.5E-5 $M_{\sun} yr^{-1}$ \\
$\rho_{1}$ & 1.74E-13 - 2.40E-13 g $cm^{-3}$ \\
$R_{env,min}$ & 7.8 $R_{sub}$ ($\sim$36 $R_{*}$)\\
$R_{env,max}$ & 4000 AU \\
$R_{c}$ & 30 AU \\
$\theta_{cav}$ & 28$\degr$ \\
$\theta_{in}$ & 53$\degr$ - 59$\degr$ \\
$M_{disk}$ & 2.6E-2 $M_{\sun}$ \\
$\dot{M}_{disk}$ & 9.54E-7 $M_{\sun} yr^{-1}$ \\
$R_{d,min}$ & 7.8 $R_{sub}$ ($\sim$36 $R_{*}$)\\
$R_{d,max}$ & 30 AU \\
$\beta$ & 1.012 \\
$\alpha$ & 2.012 \\
$L_{*}$ & 1.24 $L_{\sun}$ \\
$L_{shock}$ & 1.97 $L_{\sun}$ \\
$L_{disk}$ & 0.034 $L_{\sun}$ \\
$L_{tot}$ & 3.2 $L_{\sun}$ \\
                     
\enddata
\end{deluxetable}

Fig.~\ref{modeling} shows the best-fits obtained for our target. The main parameters that significantly effect the model SEDs are the inclination angle, the envelope density, and the centrifugal radius. As mentioned, decreasing the centrifugal radius implies that high-density material can accumulate closer to the protostar. This results in more flux longward of $\sim$20 $\micron$ and a deeper 10 $\micron$ silicate absorption feature. The flux shortward of $\sim$8 $\micron$ however decreases since decreasing $R_{c}$ increases the optical depth to the center of the envelope, due to which lesser photons can escape through the cavity walls and the upper disk layers. Another important effect is the wavelength at which the peak in far-IR emission is obtained. Increasing $R_{c}$ shifts the region where infalling material piles up farther out to a larger radius, thus shifting the far-IR peak towards longer wavelengths. The MIPS/70 $\micron$ observation is thus important in constraining the centrifugal and the outer disk radii, and a value of about 30 AU for these parameters provides a good fit to the 70 $\micron$ measurement for 2M171123. The effects of decreasing $R_{c}$ are similar to increasing the inclination angle, as both result in an increase in the amount of absorbing material in our line-of-sight. Therefore a smaller inclination angle results in larger NIR fluxes, more emission at IRAC wavelengths, a shallower silicate absorption feature, while a decrease in fluxes near $\sim$100 $\micron$. Fig.~\ref{modeling} shows that while the 56$\degr$ bin (Fit 1; covers inclinations between 53$\degr$ and 59$\degr$) provides a good fit to the full depth of the silicate feature, the MIPS data points as well as the peak in the IRS spectrum, the fluxes remain small to fit the CTIO/$K_{s}$ measurement. A better fit to this point (Fit 2) is obtained by slightly decreasing the envelope accretion rate from 2.5E-5 to 1.8E-5 $M_{\sun}$/yr, thus decreasing the envelope density. Making the envelope less dense allows more photons to escape through the cavity regions, thus increasing the NIR flux. The silicate absorption is also shallower for a less dense medium. The observed $K_{s}$-band variability could thus be explained by a variable envelope accretion rate. This mass infall rate is related to the reference density $\rho_{1}$, which is the density the envelope would have at 1AU for the limit $R_{c}$ = 0 (Kenyon et al. 1993):

\begin{equation}
\rho_{1} = 5.3 \times 10^{-14}  \left(\frac{\dot{M}_{env}}{10^{-5} M_{\sun} yr^{-1}}\right)  \left(\frac{M_{*}}{1 M_{\sun}}\right)^{-1/2}.
\end{equation}

\noindent For the small range of infall rate for 2M171123, the reference density is found to be between 2.4E-13 and 1.7E-13 $gm/cm^{3}$. This is similar to the values obtained by Kenyon et al. (1993) for Taurus protostars, but is an order of magnitude larger than that reported by Furlan et al. (2008) in the same star-forming region. We note that 2M171123 is a deeply embedded source (B07) and lies close ($\sim$0.1 pc) to the $C^{18}O$ peak that probes the densest, coldest envelope region, and so a larger envelope density is expected. 

Other parameters produce small variations to the model output. Among the disk parameters, more viscous energy is generated in the disk with an increased disk accretion rate, resulting in larger mid-IR fluxes. As the disk emission peaks near $\sim$7 $\micron$ for the viewing angle used, the parameter $\dot{M}_{disk}$ is constrained by the 5-8 $\micron$ part of the spectrum. An $\dot{M}_{disk}$ of 9.5$\times 10^{-7} M_{\sun}$/yr for 2M171123 is similar to the typical accretion rates observed among young protostars ($\sim 1 \times 10^{-8} M_{\sun}$/yr; e.g., Whitney et al. 2003). The outer disk radius was set equal to the centrifugal radius. The inner disk radius is set by the star's magnetosphere. We set it equal to a few stellar radii, similar to the inner envelope radius. Since a cavity is included, the inner envelope radius will be farther out from the central protostar. The outer envelope radius mainly effects the SED for $\lambda >$ 100 $\micron$ and can be constrained by the submillimeter and millimeter fluxes. However very large radii give too much optical depth to the center of the envelope, thus effecting the mid-IR fluxes. A value of $\sim$4000 AU provides a good fit. 

Fig.~\ref{morph2}c shows a 0.016$\times$0.016 pc model image of 2M171123, using the parameters from Fit 2. The model image shown has been obtained at an inclination of 60$\degr$, and was convolved with a Gaussian PSF of FWHM equal to that of 2M171123 in the CTIO/$K_{s}$ observations. A foreground reddening of $A_{v}$ = 25.7 (B07) has been added, and the image has been rotated so as to match the observed northeast edge of the cavity. Making the system more inclined ($i\ga$80$\degr$) results in a bipolar cavity (e.g., Fig. 8 in Whitney et al. 2003), while decreasing it to $\sim$40$\degr$ produces no outflow emission. Since reflected nebulosity is mainly observed in the northeast direction of 2M171123 and the outflow is known not to be bipolar (O99), a unipolar cavity produced at 50-60$\degr$ inclinations is a better match to the observed morphology. 

The two main heating mechanisms in the protostellar system are accretion and protostellar irradiation. During the T Tauri accretion phase, the star's magnetosphere disrupts the accretion disk at its inner edge, causing accreting material to fall along the accretion columns at velocities of the order of $\sim$200 km/s onto the star (e.g., Hartmann 2000). The hot continuum emission seen in the optical and the ultraviolet wavelengths is produced by the accretion energy dissipated when the hot gas shocks at the stellar surface. The total accretion luminosity, $L_{acc}$, is therefore a sum of the viscous accretion energy generated by the disk, $L_{disk}$, and the energy dissipated at the accretion shocks, $L_{shock}$. The total system luminosity is then

\begin{equation}
L_{tot} = L_{star} + L_{shock} + L_{disk}.
\end{equation}

Table 2 lists the luminosities from these three components. For a large inner disk radius as in the case of 2M171123 ($R_{d,min}\sim36 R_{*}$), the energy dissipated at the hot spots will be the main heating source than viscous accretion within the disk. Almost 60\% of the total energy for 2M171123 is generated at these hot spots, that cover an area of 0.01 of the stellar surface, while only $\sim$1\% is contributed by the disk. We note that a $L_{tot}$ of 3.2$L_{\sun}$ as obtained from the models is a factor of $\sim$1.5 larger than the bolometric luminosity of 2.2$L_{\sun}$ reported by B07. Similar discrepancies have been noted by Furlan et al. (2008) in their modeling of Taurus protostars, and have been explained by the inclination angle of the system. Since the contribution from disk luminosity is minimum for lowly inclined disks, model luminosity will be smaller than the bolometric luminosity. As the contribution from this component increases with higher inclinations, models may overestimate the true bolometric luminosities (e.g., Whitney et al. 2003). 

None of the two models (Fits 1 and 2) at a small aperture of about 200 AU are able to reproduce the submillimeter and millimeter measurements, and a much larger aperture of 5200 AU is required to fit these two points (Fit 3). The absence of any one good model that would fit these points along with the shorter-wavelength data indicates that the long-wavelength emission arises from a cold dust component that lies outside of the infall region. The 350$\micron$ measurement corresponds to the flux integrated over a beam 40$\arcsec$ in diameter (Wu et al. 2006), while the 1.3 mm observation is obtained from a 22$\arcsec$ diameter aperture (Reipurth et al. 1996). We note that the millimeter observation plotted is actually of the source B59-MMSI that lies $\sim$15$\arcsec$ from 2M171123, and therefore must include more emission from the surrounding dense molecular cloud than the protostellar system itself. The large aperture of 40$\arcsec$ for the 350 $\micron$ measurement also indicates that cold dust from the surrounding dense cloud contributes a larger fraction to the observed emission at this wavelength.

\section{Thickness of Ice Mantle}
\label{ices}

Recent surveys with {\it Spitzer}/IRS of a large number of protostellar systems have revealed a suite of absorption features, the most prominent being the ice features at 6.0 and 6.8 $\micron$, the 10 $\micron$ feature due to silicates, and the 15 $\micron$ $CO_{2}$ ice band (e.g., Boogert et al. 2008, Kessler-Silacci et al. 2005). At the low temperatures that exist in such dense molecular clouds and circumprotostellar envelopes, atoms and molecules freeze out on dust grains, creating ice mantles that remain preserved in such cold environments. Thermal processes such as stellar irradiation or shocks related to accretion activity or outflows result in a thinning of these ice mantles, and alter the composition of the ice particles (e.g., Boogert et al. 2008).

\begin{figure}
\plottwo{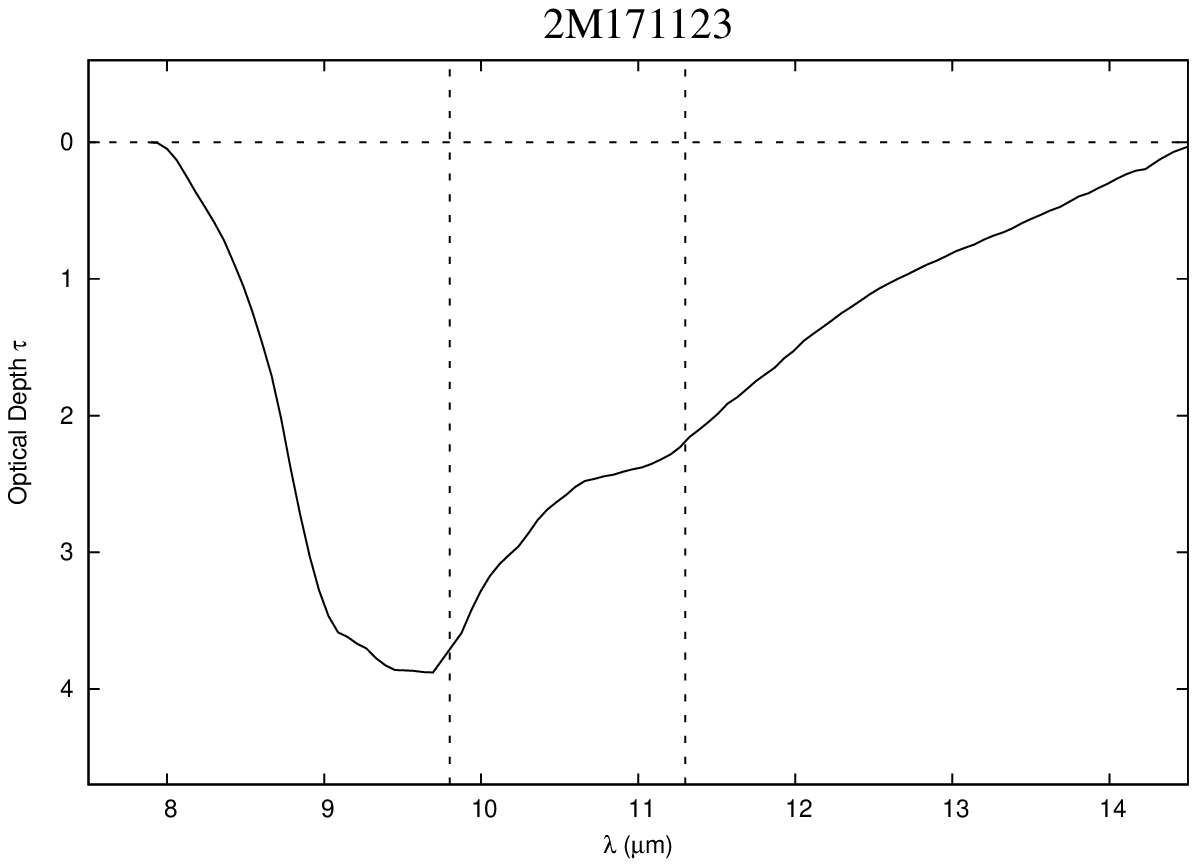}{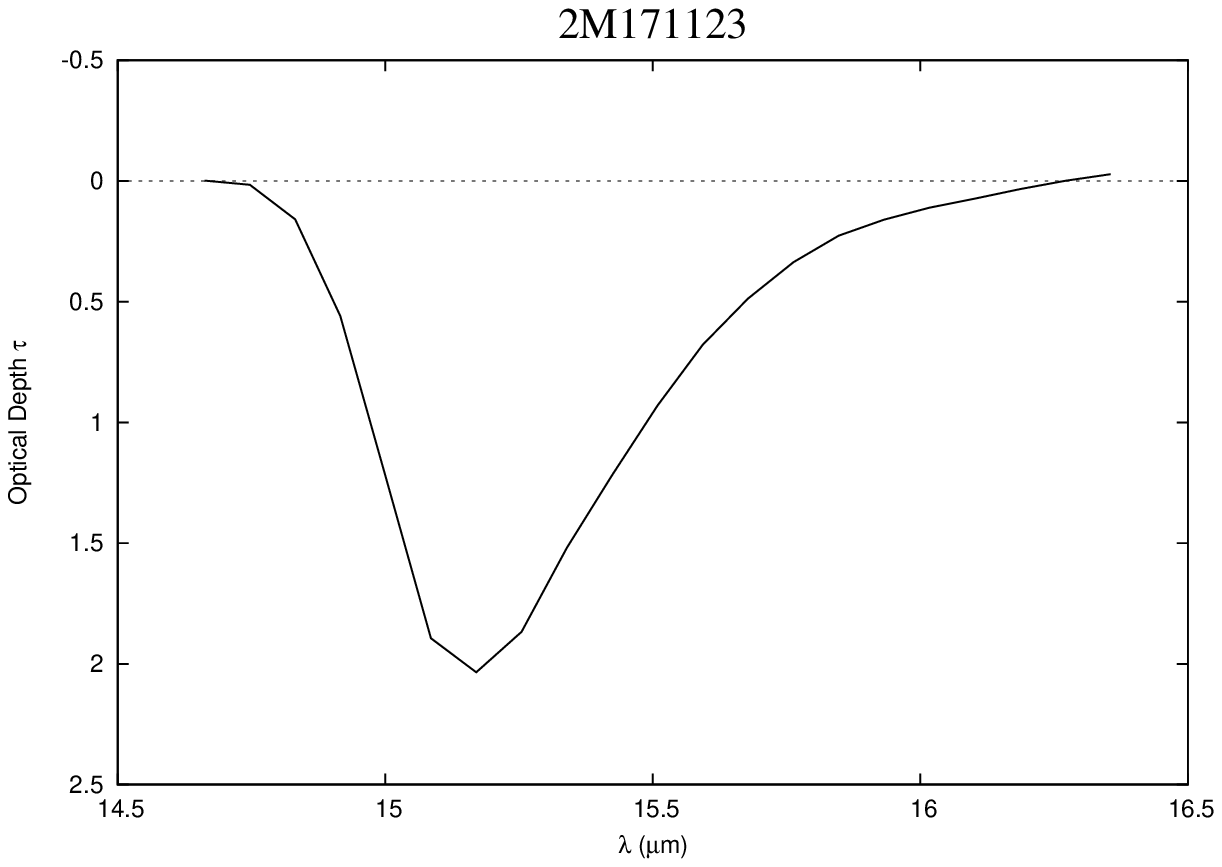}
\plottwo{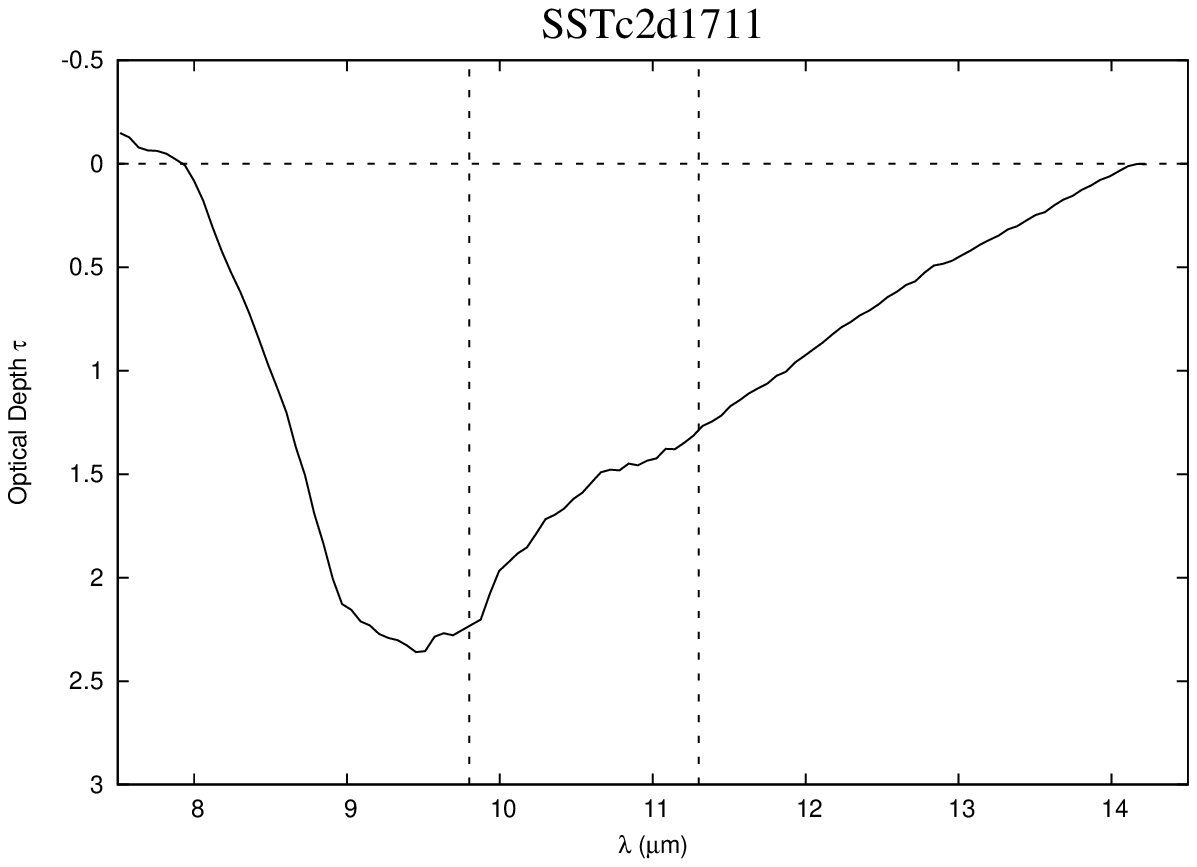}{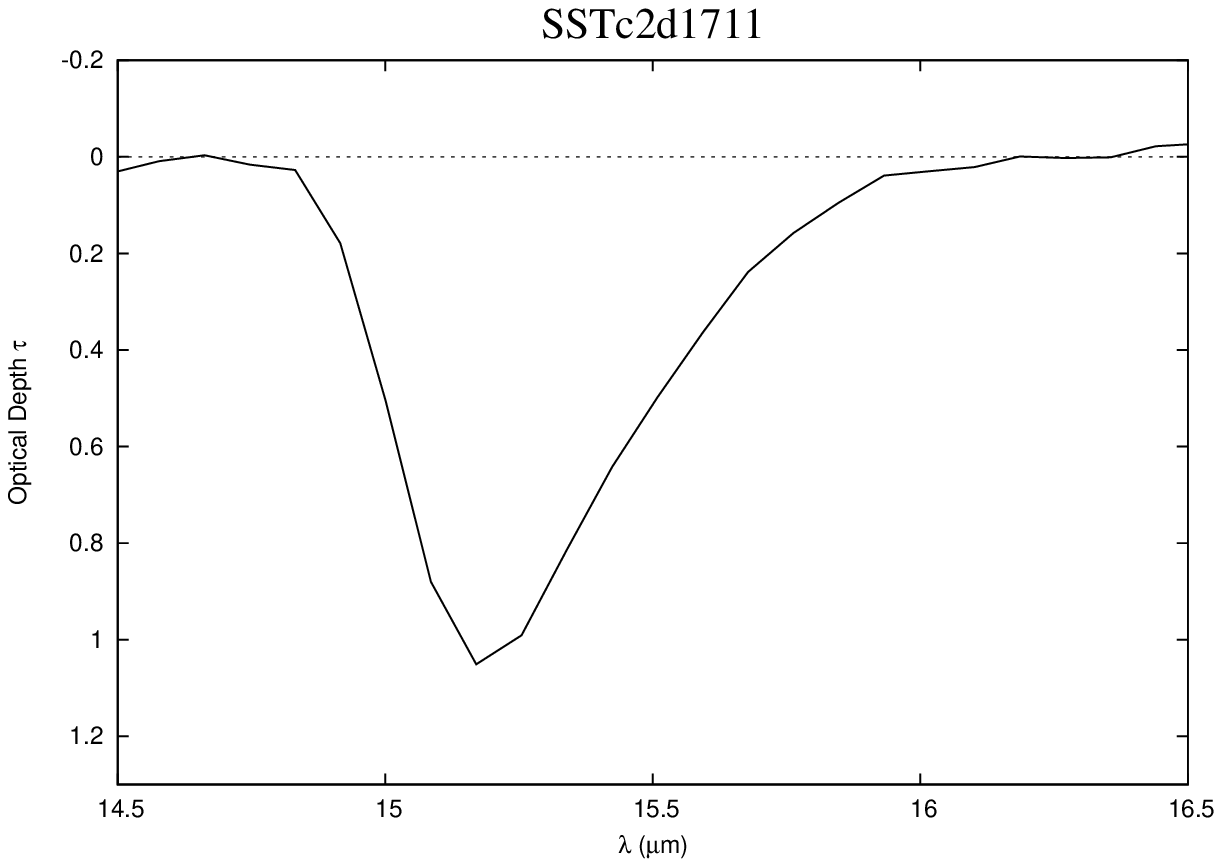}
\plottwo{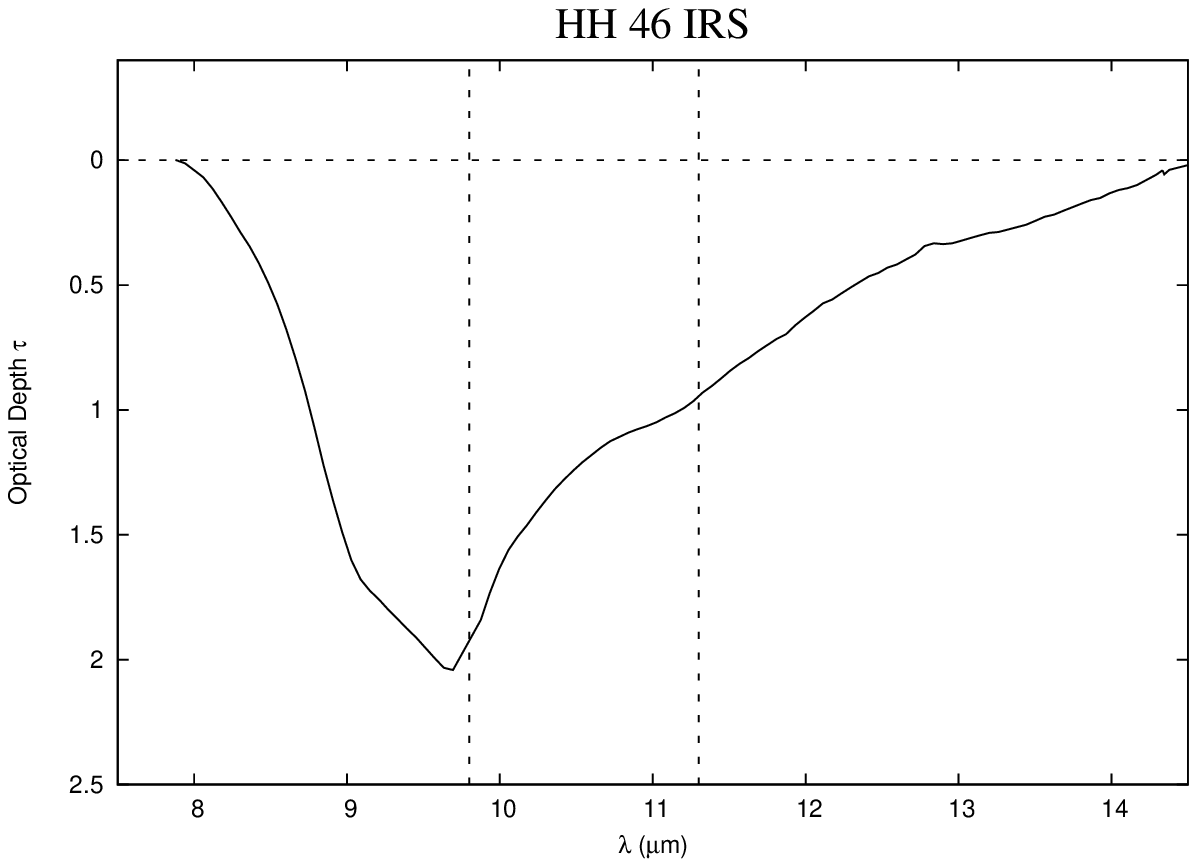}{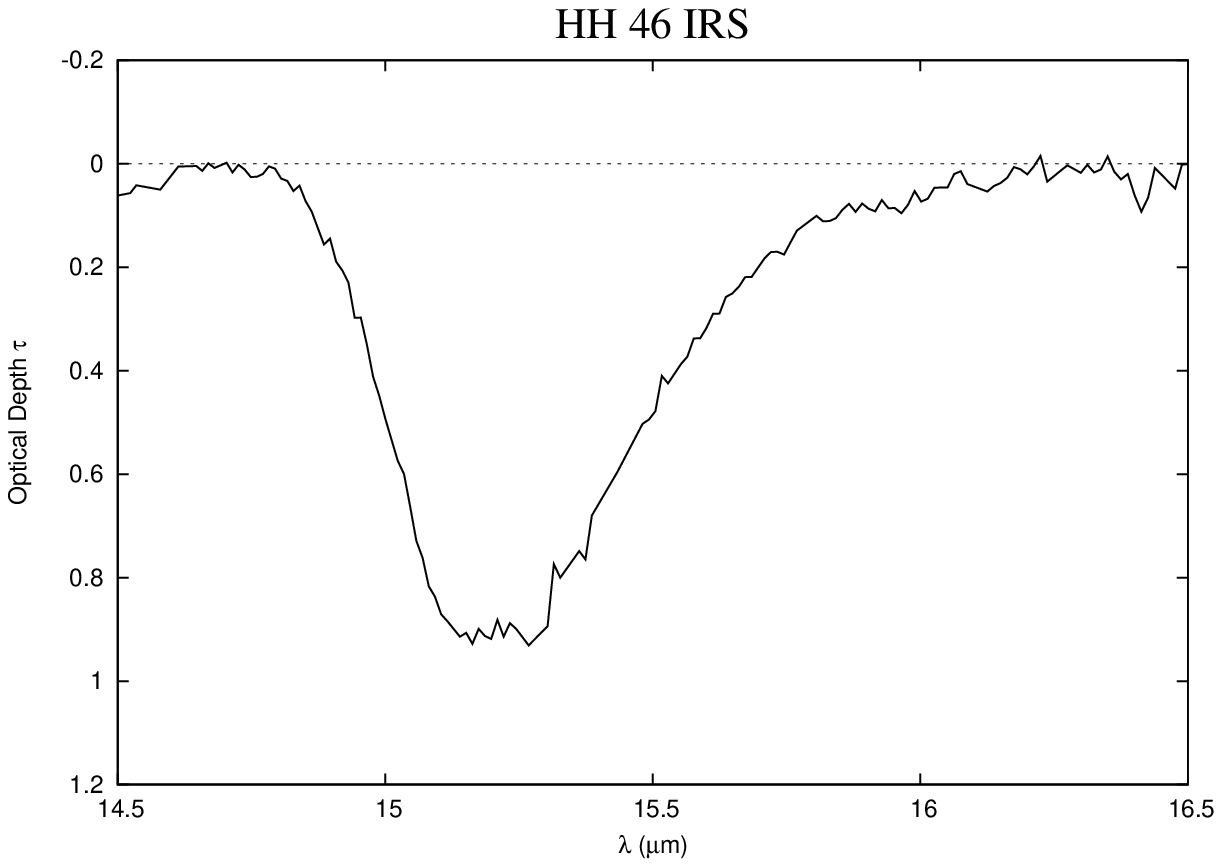}
\plottwo{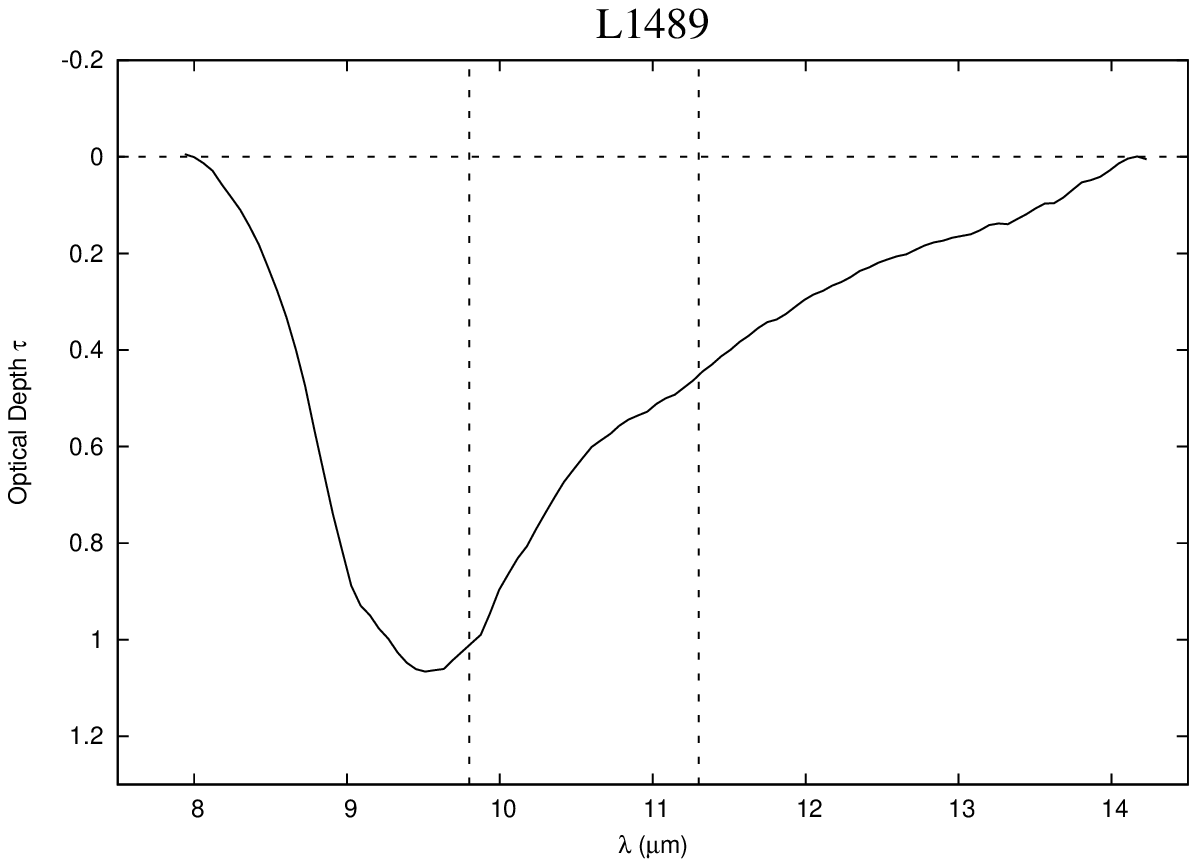}{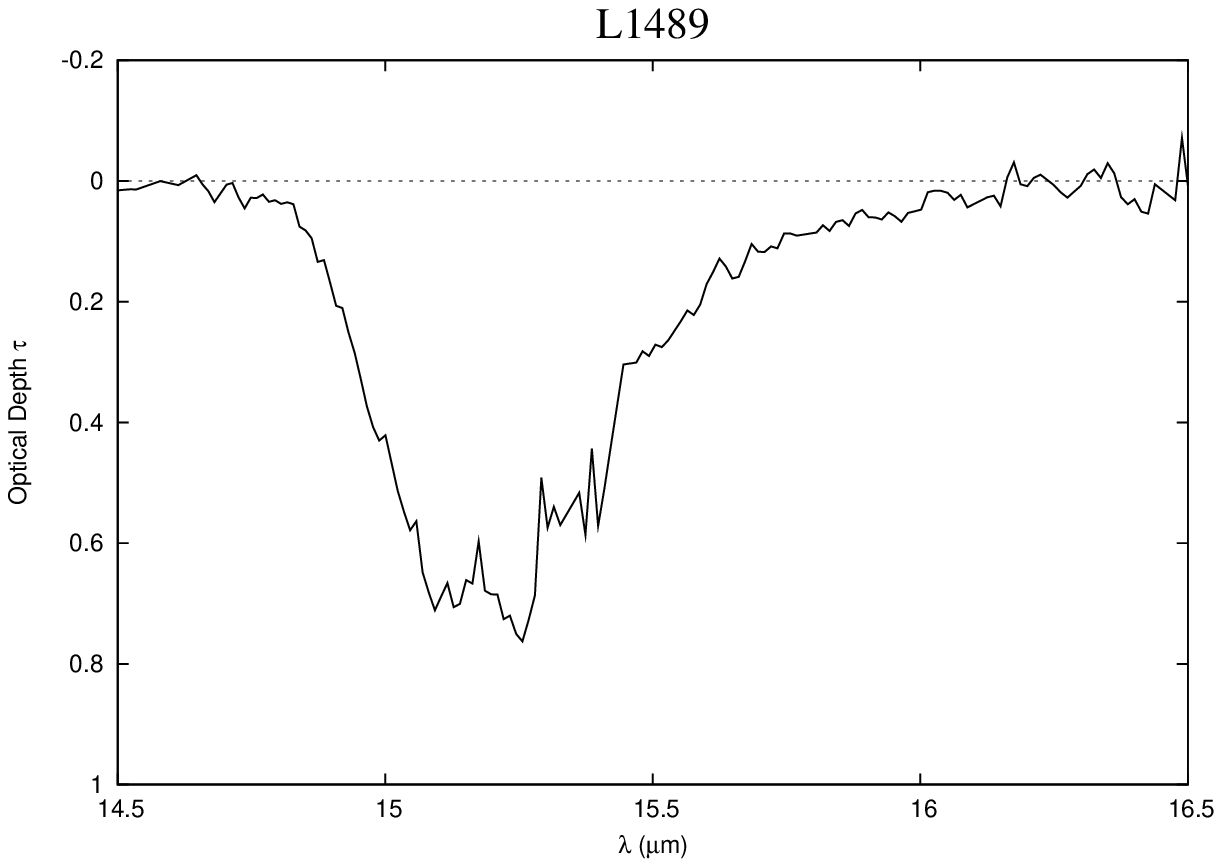}
\plottwo{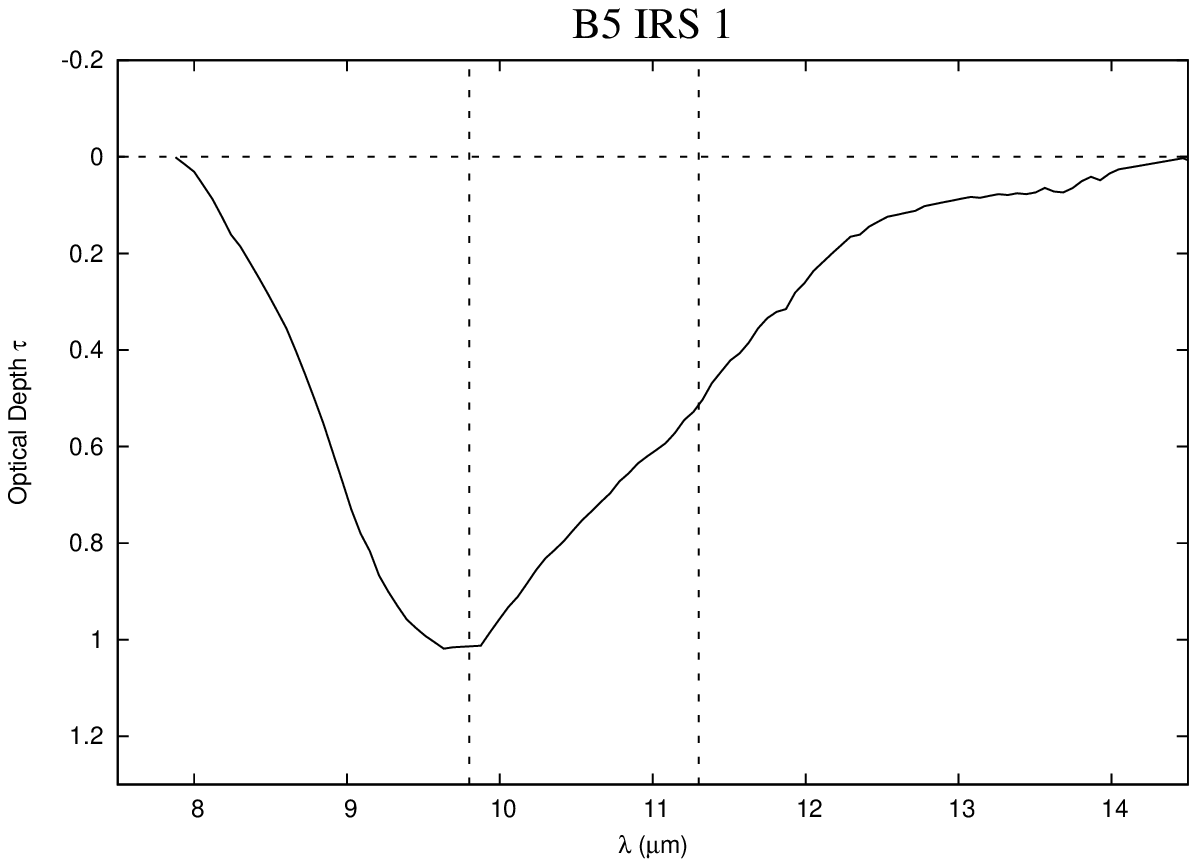}{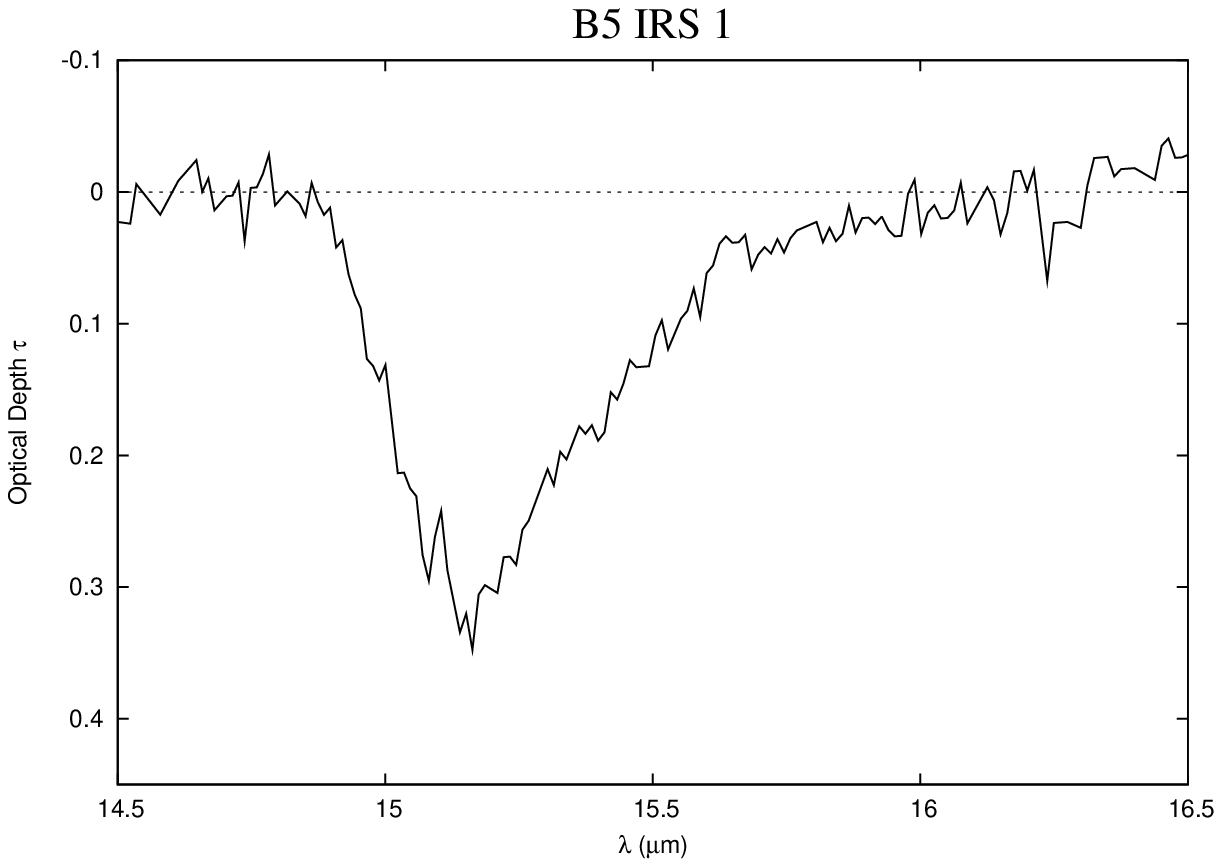}
    \caption{Optical depth in the 10 $\micron$ silicate feature ({\it left}) and the 15 $\micron$ $CO_{2}$ feature ({\it right}). From top to bottom: 2M171123, SST1711, HH 46, L1489 and B5 IRS 1.}
    \label{ices1}
 \end{figure}

Fig. \ref{ices1} shows the optical depth, $\tau$, for the 10 $\micron$ silicate and the 15 $\micron$ $CO_{2}$ absorption features for 2M171123. To measure the optical depth, the continuum in the 10 $\micron$ silicate feature was  approximated by a straight line that connects the 8 and 13 $\micron$ fluxes (e.g., Kessler-Silacci et al. 2005), while in the 15 $\micron$ feature it was approximated with a straight line connecting the 14.65 and 16.3 $\micron$ fluxes (e.g., Quanz et al. 2007). The optical depth was then computed using

\begin{equation}
F_{\nu}/F_{c} = e^{-\tau},
\end{equation}

\noindent where $F_{\nu}$ and $F_{c}$ are the observed and continuum fluxes, respectively. The silicate spectra for Class I objects are known to exhibit smooth, featureless absorption profiles with a strong, narrow peak at 9.6$\micron$, indicative of amorphous silicates (Kessler-Silacci et al. 2005). Stronger absorption is observed in the more embedded protostars. A noticeable feature in the 2M171123 silicate spectrum is the significantly lower optical depth between $\sim$9.8  and 11.3 $\micron$. Such rare shallow absorption features have been reported previously for some low-mass embedded protostars, such as HH 46 IRS (IRAS 08242-5050), B5 IRS 1 (IRAS 03445+3242) (Boogert et al. 2004) and L1489 (IRAS 04016+2610) (Kessler-Silacci et al. 2005). This `shoulder' near 11.3 $\micron$ due to enhanced emission at this wavelength may be related to the presence of crystalline forsterites, commonly observed in older, more evolved Class II systems. The transition from amorphous to crystalline silicates results in a shift in the wavelength of maximum emission from 9.7 to 11.3 $\micron$ (e.g., Kessler-Silacci et al. 2005). Such transitions however occur on a timescale longer than the age of protostellar systems, which is why the presence of crystalline silicates in protostars seems improbable. The main formation mechanism of crystalline grains in T Tauri disks is considered to be thermal annealing in the warm inner disk regions, at temperatures $>$1000 K. Some low-temperature crystallization processes (e.g., Molster et al. 2001) may occur in these cold embedded sources, that may result in early crystallization of silicates present in the envelope. This would however suggest crystalline silicates to be ubiquitous among all protostars, and it is not clear why such processes may be active in just a few of these systems. The presence of a heating mechanism such as an outflow also could not have expedited silicate crystallization, since 2M171123, HH 46, B5 IRS 1 and L1489 (IRAS 04016+2610) are all associated with an outflow source (Langer et al. 1996; Noriega-Crespo et al. 2004; Gomez et al. 1997), but display varying strengths in the 11.3 $\micron$ feature (Fig. \ref{ices1}).

To understand this feature better, we have compared the 10 and 15 $\micron$ spectra for 2M171123 with a few other low-mass protostars, including B5 IRS 1, HH 46 and L1489. Among these, SSTc2d J171122.2-272602 (hereafter SST1711) is another Class 0 object identified in B59, and lies close to the CS molecular gas peak, as well as the extinction peak (B07). The 15 $\micron$ $CO_{2}$ bending mode is an important tracer of ice composition, and the extent of ice processing that has occurred in the system due to various heating effects (e.g., Boogert et al. 2002). 2M171123 and SST1711 show smooth, featureless $CO_{2}$ profiles, similar to the spectrum of the background star CK 2 that probes the ice composition of the dark molecular cloud material, and is thus representative of unprocessed ice (Knez et al. 2005). HH 46 and L1489 show a double-peaked substructure at the base of the $CO_{2}$ band that is caused by crystallization and effective segregation of the $H_{2}O$ and $CO_{2}$ ice species (e.g., Boogert et al. 2004). The $H_{2}O$ crystallization requires a temperature of at least 50K, while sublimation temperatures are higher ($\sim$90K; Boogert et al. 2002). The transition from amorphous to crystalline ice thus occurs in this temperature range, suggesting that at least some fraction of the inner envelope for HH 46 and L1489 is warm enough ($\ga$ 50K) to heat up the ice and cause the observed substructures (e.g., Boogert et al. 2004). The smooth $CO_{2}$ profiles for 2M171123 and SST1711 thus indicate temperatures below $\sim$50K in their circumprotostellar envelopes and surrounding environment.  

An inverse trend is noticeable in Fig.~\ref{ices1} between the strength in the 11.3 $\micron$ shoulder and the optical depth in the $CO_{2}$ band, with larger optical depths observed between 9.8 and 11.3 $\micron$ for sources with deeper absorption at 15 $\micron$. The origin of the 11.3 $\micron$ shoulder could be related to the thickness of the ice mantle coated on the silicate grains. As noted by Li et al. (2008), the inclusion of an ice mantle on silicate dust produces a shoulder at $\lambda$ $>$ 11 $\micron$, that may be interpreted as a crystalline silicate feature. A thicker ice mantle would then result in a more prominent shoulder at this wavelength. 

\begin{figure}
\plottwo{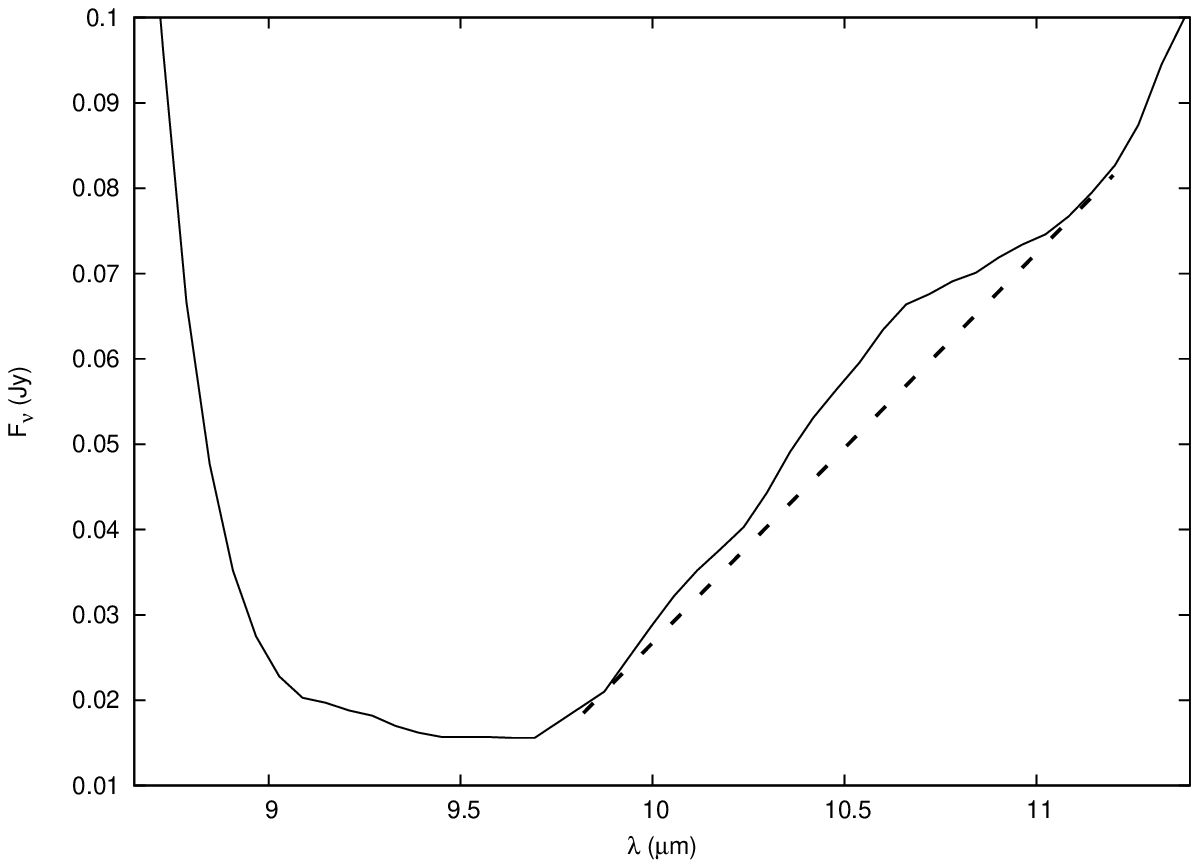}{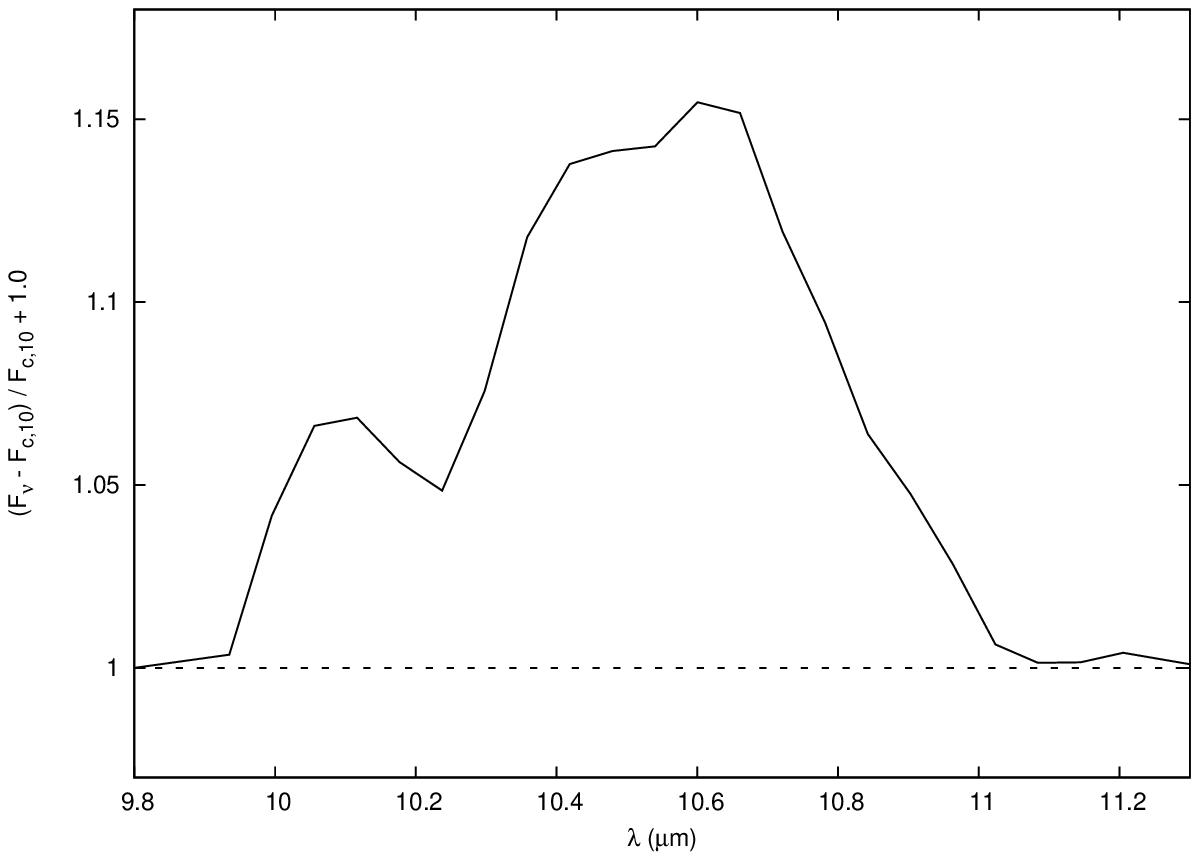}
\plottwo{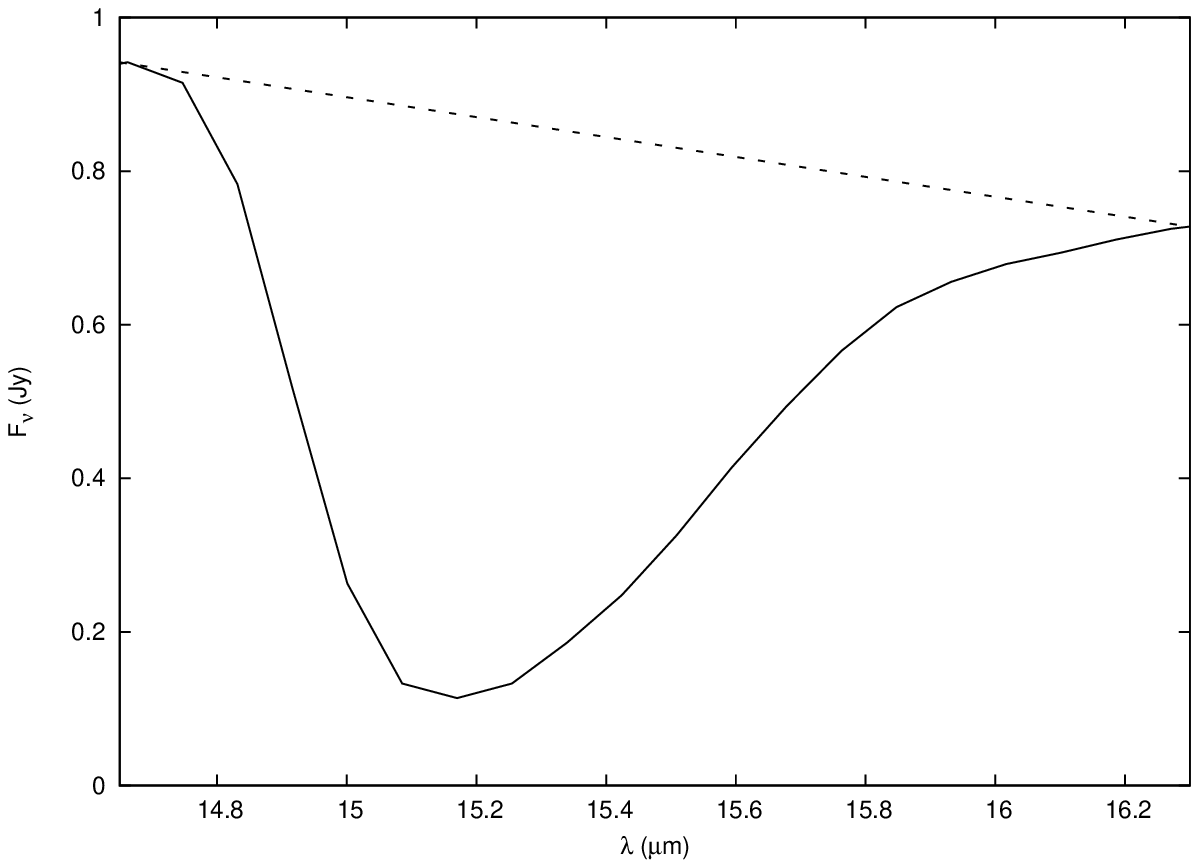}{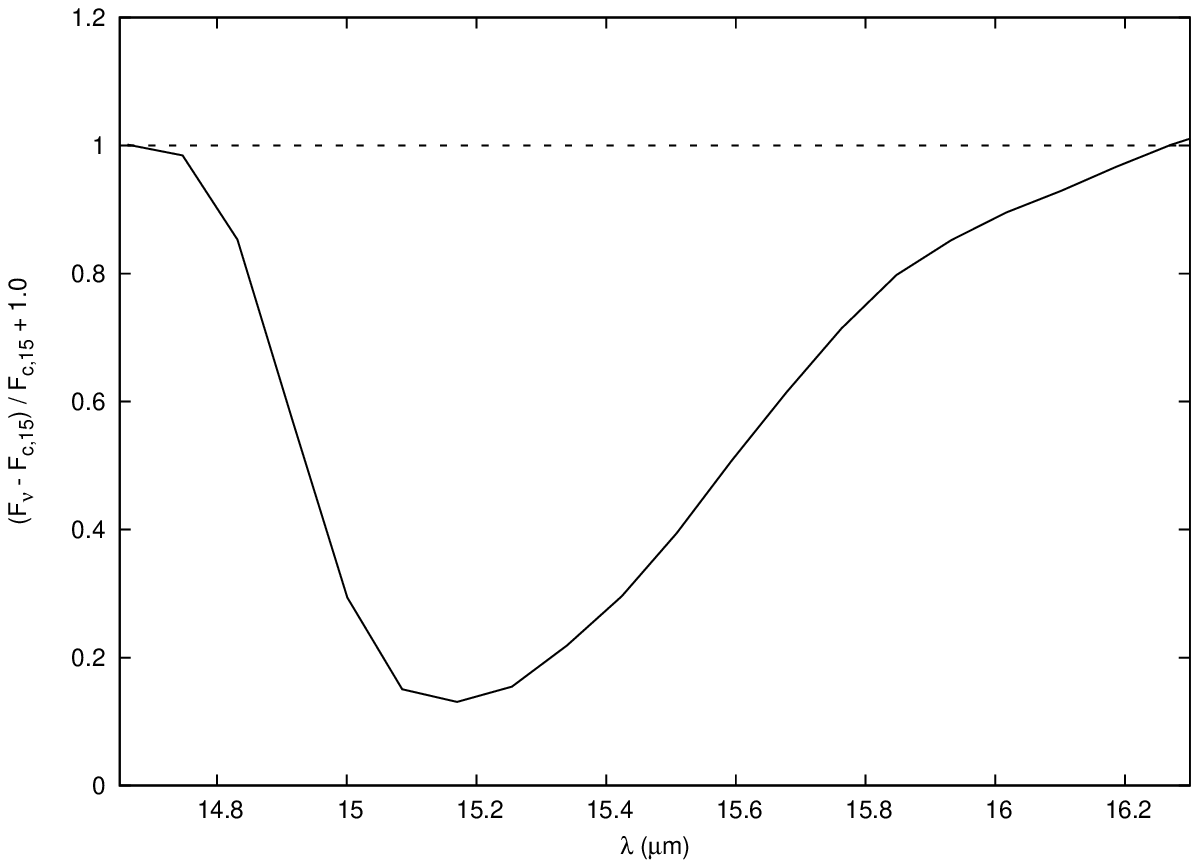}
    \caption{{\it Top, left} (a): The observed 8-13 $\micron$ silicate spectrum for 2M171123. Dotted line marks the  continuum flux $F_{c,10}$ approximated for the 9.8-11.3 $\micron$ shoulder. {\it Top, right} (b): The normalized continuum-subtracted fluxes in the 9.8-11.3 $\micron$ region. {\it Bottom, left} (c): The observed fluxes in the 15 $\micron$ $CO_{2}$ ice-band for 2M171123. Dotted line marks the continuum flux $F_{c,15}$ approximated in the 14.65-16.3 $\micron$ wavelength range. {\it Bottom, right} (d): The normalized continuum-subtracted fluxes in the 15 $\micron$ feature.}
    \label{crys}
 \end{figure}

To confirm this hypothesis, we measured the strength in this feature, $F_{peak}$, and checked for any possible correlations with the water-ice column density. To measure $F_{peak}$, we first approximated the continumm below the feature by a straight line that connects the observed fluxes at 9.8 and 11.3 $\micron$, as shown in Fig.~\ref{crys}a (dotted line) for the case of 2M171123. These particular end points were selected since the shoulder is observed in this narrow wavelength range. The feature strength was then defined as:

\begin{equation}
F_{peak}=maximum ~ [(F_{\nu} - F_{c,10})/F_{c,10} + 1.0],
\end{equation}

\noindent where $F_{c,10}$ is the underlying continuum in the 9.8-11.3 $\micron$ range, and ($F_{\nu} - F_{c,10})/F_{c,10}$ are the normalized continuum-subtracted fluxes, plotted in Fig.~\ref{crys}b. Thus a larger value for $F_{peak}$ would imply a greater contrast above the continuum, and therefore a more prominent shoulder would be observed. We repeated the same analysis for the 15 $\micron$ $CO_{2}$ absorption feature. In this case, the {\it overlying} continuum, $F_{c,15}$, was approximated by connecting the observed fluxes at 14.65 and 16.3 $\micron$, as shown in Fig.~\ref{crys}c. The normalized continuum-subtracted fluxes for the $CO_{2}$ feature are plotted in Fig.~\ref{crys}d. We define a parameter, $F_{15.2}$, as a measure of the contrast {\it below} the continuum $F_{c,15}$, i.e. a smaller value for $F_{15.2}$ would imply deeper absorption in this ice feature. Using this method, we obtained $F_{peak}$ and $F_{15.2}$ values for a set of low-mass protostellar systems, including 2M171123 and other objects shown in Fig.~\ref{ices1}. These systems were selected based on the availability of the ice column density measurements, which have been obtained from Boogert et al. (2008).

\begin{figure}
\plotone{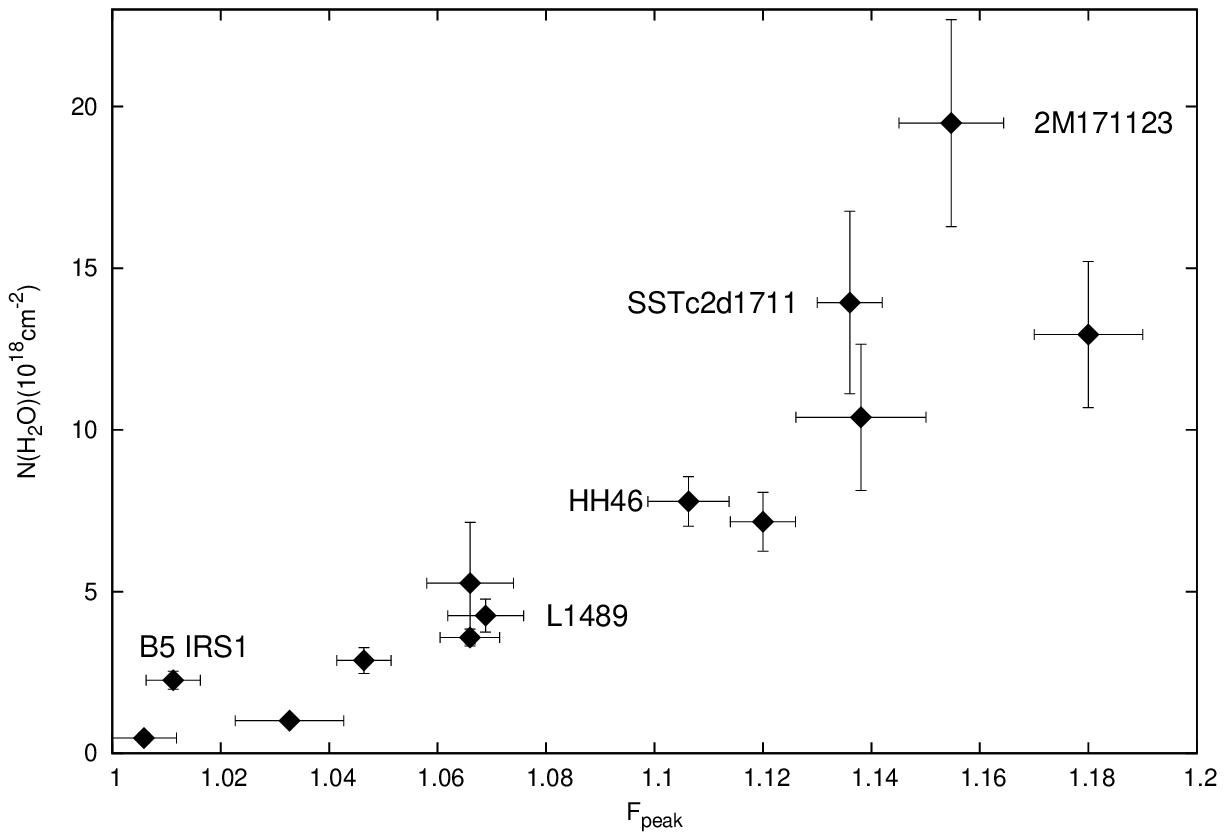}
\plotone{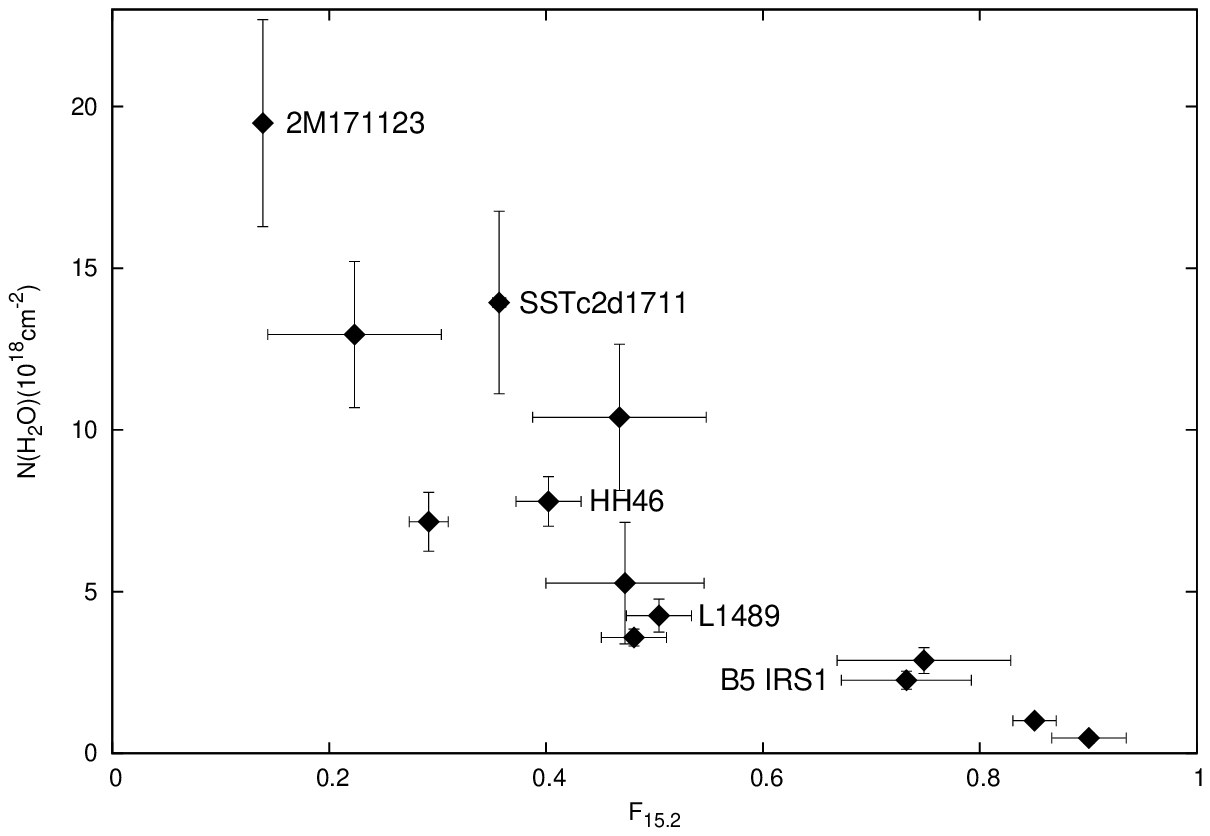}
\plotone{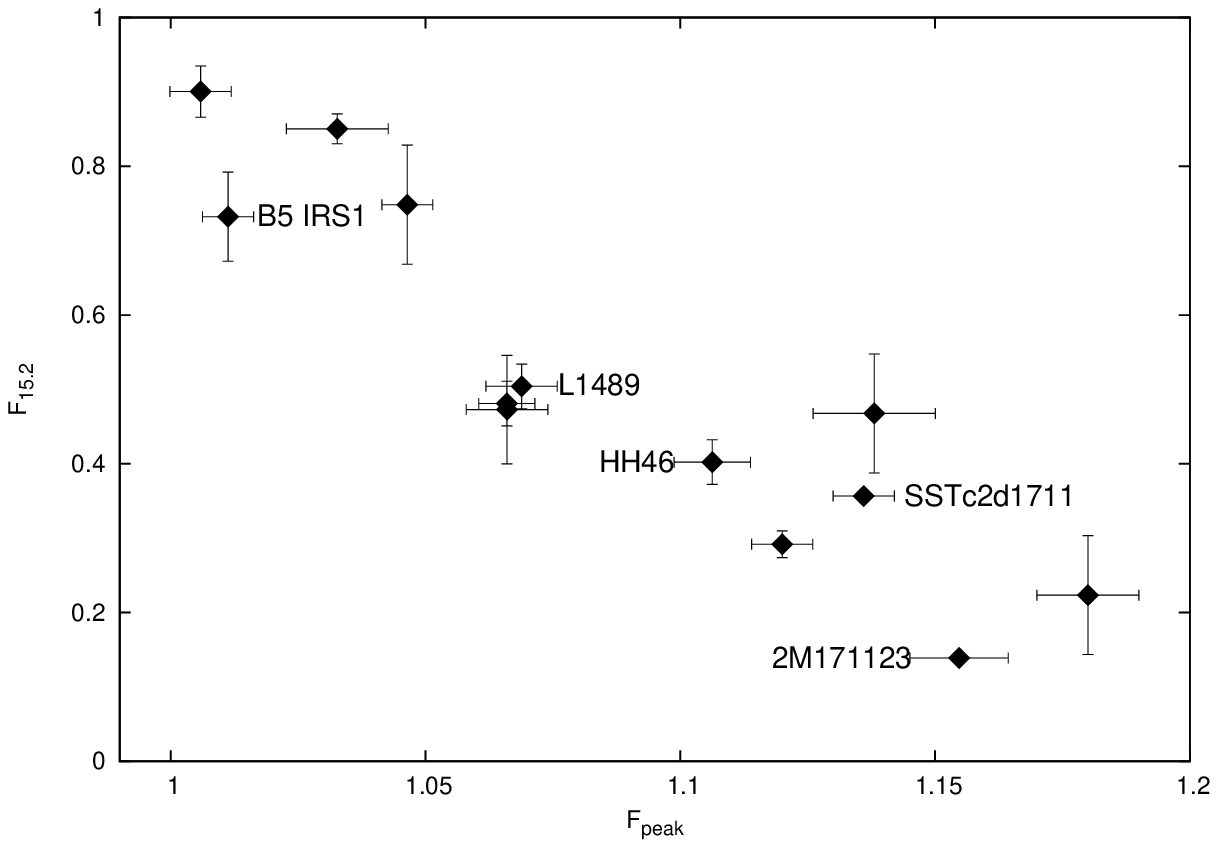}
    \caption{{\it Top, (a)}: $F_{peak}$ versus the $H_{2}O$-ice column density. {\it Middle, (b)}: $F_{15.2}$ versus the ice column density. {\it Bottom, (c)}: $F_{peak}$ versus $F_{15.2}$. The thicker the ice mantle over silicate grains, the more prominent is the 11.3 $\micron$ feature. }
    \label{ices2}
 \end{figure}

Fig. \ref{ices2}a plots $F_{peak}$ versus the water-ice column density, $N(H_{2}O)$. We note that both $F_{peak}$ and $F_{15.2}$ have values in units of $(F_{\nu} - F_{c,10})/F_{c,10} + 1.0$. An exponentially-increasing trend is observed between the two parameters, with $F_{peak}$ increasing towards thicker ice mantles. $F_{15.2}$ shows an exponentially decreasing trend with $N(H_{2}O)$ (Fig. \ref{ices2}b), as expected since a larger value for this parameter indicates shallower absorption in the 15 $\micron$ ice band, and correlates with a smaller ice column density. In Fig. \ref{ices2}c, $F_{peak}$ is found to increase linearly with increasing depth in the ice feature, further confirming that objects with deeper ice absorption show greater strengths in the 11.3 $\micron$ shoulder. The lack of crystallization signatures in the $CO_{2}$ absorption band for 2M171123 thus correlates well with the presence of a thick ice mantle and a more prominent 11.3 $\micron$ edge. Thus protostars that are not deeply embedded or have heated their envelopes sufficiently to melt away a large fraction of the ice coatings exhibit a weaker strength in the 11.3 $\micron$ shoulder, confirming the origin of this feature in the thickness of ice coating over the grains. We note that some crystalline silicates may be present in the circumprotostellar envelopes, as ISM-like dust is known to show a 3-5\% crystalline mass fraction (e.g., Li et al. 2008). Nevertheless, it is the thickness of the ice coating over the silicates grains that seems to effect the observed 11.3 $\micron$ shoulder strongly.

\section{Conclusions}
We present NIR observations of the low-mass protostar 2M171123 in the B59 molecular cloud. Scattered light nebulosity possibly tracing an outflow cavity are observed towards this source. Modeling this system indicates slight variability in the mass infall rate between 2.5E-5 and 1.8E-5 $M_{\sun}$/yr, that could explain the observed $K_{s}$-band variability. The system is viewed at an intermediate inclination of 56$\degr$ to the line-of-sight. We report the detection of a low-luminosity Class I object 2M17112255, that lies $\sim$8$\arcsec$ from 2M171123. We estimate a mass of $\sim$0.12-0.25 $M_{\sun}$ for this source, at an age of 0.1-1Myr. 2M171123 shows a smooth, featureless ice profile with no signs of ice processing, indicating a circumprotostellar envelope at a temperature below $\sim$50 K. We find a strong correlation between the thickness of ice coating over the silicate grains, and the strength in the rarely observed 11.3 $\micron$ absorption feature. The enhanced 11.3 $\micron$ emission observed towards 2M171123 must be due to a thicker ice mantle over the silicate grains.

\acknowledgments
We wish to thank Thomas Robitaille and Barbara Whitney for helpful comments and suggestions regarding modeling of the system. B. R. would like to thank M. R. Zapatero Osorio for a discussion on evolutionary models. Support for this project has been provided by CONSTELLATION grant \# YA 2007, and the Spanish
Ministry of Science via project AYA2007-67458. This work is based on observations obtained at the Cerro Tololo Inter-American Observatory, National Optical Astronomy Observatory, which are operated by the Association of Universities for Research in Astronomy, under contract with the National Science Foundation. This work is based in part on observations made with the {\it Spitzer Space Telescope}, which is operated by the Jet Propulsion Laboratory, California Institute of Technology under a contract with NASA. Support for this work was provided by NASA through an award issued by JPL/Caltech. This publication makes use of data products from the Two Micron All Sky Survey, which is a joint project of the University of Massachusetts and the Infrared Processing and Analysis Center/California Institute of Technology, funded by the National Aeronautics and Space Administration and the National Science Foundation.

\end{document}